\documentclass{emulateapj}
\usepackage{epsfig,natbib}
\usepackage{graphicx}
\newcommand{\ms}{\mbox{m\,s$^{-1}~$}}

\newcommand{\msun}{M$_{\odot}~$}
\newcommand{\msune}{M$_{\odot}$}

\newcommand{\rsune}{R$_{\odot}$}

\newcommand{\mearth}{$M_\earth$~}
\newcommand{\mearthe}{$M_\earth$}
\newcommand{\rearth}{$R_\earth$~}
\newcommand{\rearthe}{$R_\earth$}

\newcommand{\feh}{\ensuremath{[\mbox{Fe}/\mbox{H}]}}

\newcommand{\sini}{\ensuremath{\sin i}}
\newcommand{\msini}{\ensuremath{M_{\rm p} \sin i}}

\newcommand{\teff}{\ensuremath{T_{\rm eff}}}

\newcommand{\logg}{\ensuremath{\log{g}}}

\newcommand{\kepler}{\emph{Kepler}~}
\newcommand{\Kepler}{\emph{Kepler}~}
\newcommand{\Keplere}{\emph{Kepler}}

\newcommand{\kp}{\ensuremath{\mathrm{Kp}}}
\newcommand{\rstar}{\ensuremath{R_\star}}
\newcommand{\mstar}{\ensuremath{M_\star}}

\newcommand{\rp}{\ensuremath{R_{\rm p}}}

\newcommand{\Nst}{58,041}                 
\newcommand{\Npl}{438}                      
\newcommand{\NplUniq}{375}            
\newcommand{\BinsLogArea}{28.5}     
\newcommand{\kRPL}{\ensuremath{2.9^{+0.5}_{-0.4}}}             
\newcommand{\alphaRPL}{\ensuremath{-1.92 \pm 0.11}}     
\newcommand{\kPPLRtwofour}{\ensuremath{0.064 \pm 0.040}}             
\newcommand{\betaPPLRtwofour}{\ensuremath{0.27 \pm 0.27}}             
\newcommand{\PzeroPPLRtwofour}{\ensuremath{7.0 \pm 1.9}}             
\newcommand{\gammaPPLRtwofour}{\ensuremath{2.6 \pm 0.3}}             
\newcommand{\kPPLRfoureight}{\ensuremath{0.0020 \pm 0.0012}}             
\newcommand{\betaPPLRfoureight}{\ensuremath{0.79 \pm 0.50}}             
\newcommand{\PzeroPPLRfoureight}{\ensuremath{2.2 \pm 1.0}}             
\newcommand{\gammaPPLRfoureight}{\ensuremath{4.0 \pm 1.2}}             
\newcommand{\kPPLReightthirtytwo}{\ensuremath{0.0025 \pm 0.0015}}             
\newcommand{\betaPPLReightthirtytwo}{\ensuremath{0.37 \pm 0.35}}             
\newcommand{\PzeroPPLReightthirtytwo}{\ensuremath{1.7 \pm 0.7}}             
\newcommand{\gammaPPLReightthirtytwo}{\ensuremath{4.1 \pm 2.5}}             
\newcommand{\kPPLRtwothirtytwo}{\ensuremath{0.035 \pm 0.023}}             
\newcommand{\betaPPLRtwothirtytwo}{\ensuremath{0.52 \pm 0.25}}             
\newcommand{\PzeroPPLRtwothirtytwo}{\ensuremath{4.8 \pm 1.6}}             
\newcommand{\gammaPPLRtwothirtytwo}{\ensuremath{2.4\pm 0.3}}             
\newcommand{\TeffEtaNaught}{\ensuremath{0.165 \pm 0.011}}             
\newcommand{\TeffGammaT}{\ensuremath{-0.081 \pm 0.011}}        
\newcommand{\TeffChiSqReduced}{\ensuremath{1.03}}        
\newcommand{\OccPtenRtwofour}{\ensuremath{0.025 \pm 0.003}}  
\newcommand{\OccPfiftyRtwofour}{\ensuremath{0.130 \pm 0.008}}  
\newcommand{\OccPtenRfoureight}{\ensuremath{0.005 \pm 0.001}}  
\newcommand{\OccPfiftyRfoureight}{\ensuremath{0.023 \pm 0.003}}  
\newcommand{\OccPtenReightthirtytwo}{\ensuremath{0.004 \pm 0.001}}  
\newcommand{\OccPfiftyReightthirtytwo}{\ensuremath{0.013 \pm 0.002}}  
\newcommand{\OccPtenRtwothirtytwo}{\ensuremath{0.034 \pm 0.003}}  
\newcommand{\OccPfiftyRtwothirtytwo}{\ensuremath{0.165 \pm 0.008}}  
\newcommand{\OccPtenReightthirtytwoMag}{\ensuremath{0.005 \pm 0.001}}  
\newcommand{\OccPtenRfivethirtytwoMag}{\ensuremath{0.0076 \pm 0.0013}}  

\shortauthors{Howard {et~al.}}
\shorttitle{Planet Occurrence from \Kepler}
\submitted{Submitted to ApJ --- 2011/03/13}
\begin{document}
\pagenumbering{arabic}

\title{Planet Occurrence within 0.25 AU of Solar-type Stars from \textit{Kepler}$^{\dagger}$}
\author{
Andrew~W.~Howard\altaffilmark{1,*},  
Geoffrey~W.~Marcy\altaffilmark{1},  
Stephen~T.~Bryson\altaffilmark{2}, 
Jon~M.~Jenkins\altaffilmark{3}, 
Jason~F.~Rowe\altaffilmark{2}, 
Natalie~M.~Batalha\altaffilmark{4}, 
William~J.~Borucki\altaffilmark{2}, 
David~G.~Koch\altaffilmark{2}, 
Edward~W.~Dunham\altaffilmark{5}, 
Thomas~N.~Gautier III\altaffilmark{6}, 
Jeffrey~Van Cleve\altaffilmark{3}, 
William~D.~Cochran\altaffilmark{7}, 
David~W.~Latham\altaffilmark{8}, 
Jack~J.~Lissauer\altaffilmark{2}, 
Guillermo~Torres\altaffilmark{8}, 
Timothy~M.~Brown\altaffilmark{9}, 
Ronald~L.~Gilliland\altaffilmark{10}, 
Lars~A.~Buchhave\altaffilmark{11}, 
Douglas~A.~Caldwell\altaffilmark{3}, 
J{\o}rgen~Christensen-Dalsgaard\altaffilmark{12,13}, 
David~Ciardi\altaffilmark{14}, 
Francois~Fressin\altaffilmark{8}, 
Michael~R.~Haas\altaffilmark{2}, 
Steve~B.~Howell\altaffilmark{15}, 
Hans~Kjeldsen\altaffilmark{12}, 
Sara~Seager\altaffilmark{16}, 
Leslie~Rogers\altaffilmark{16}, 
Dimitar~D.~Sasselov\altaffilmark{8}, 
Jason~H.~Steffen\altaffilmark{17}, 
Gibor~S.~Basri\altaffilmark{1}, 
David~Charbonneau\altaffilmark{8}, 
Jessie~Christiansen\altaffilmark{2}, 
Bruce~Clarke\altaffilmark{2}, 
Andrea~Dupree\altaffilmark{8}, 
Daniel~C.~Fabrycky\altaffilmark{18}, 
Debra~A.~Fischer\altaffilmark{19}, 
Eric~B.~Ford\altaffilmark{20}, 
Jonathan~J.~Fortney\altaffilmark{18}, 
Jill~Tarter\altaffilmark{3}, 
Forrest~R.~Girouard\altaffilmark{21}, 
Matthew~J.~Holman\altaffilmark{8}, 
John~Asher~Johnson\altaffilmark{22}, 
Todd~C.~Klaus\altaffilmark{21}, 
Pavel~Machalek\altaffilmark{3}, 
Althea~V.~Moorhead\altaffilmark{20}, 
Robert~C.~Morehead\altaffilmark{20}, 
Darin~Ragozzine\altaffilmark{8}, 
Peter~Tenenbaum\altaffilmark{3}, 
Joseph~D.~Twicken\altaffilmark{3}, 
Samuel~N.~Quinn\altaffilmark{8}, 
Howard~Isaacson\altaffilmark{1},  
Avi~Shporer\altaffilmark{9,23}, 
Philip~W.~Lucas\altaffilmark{24}, 
Lucianne~M.~Walkowicz\altaffilmark{1},  
William~F.~Welsh\altaffilmark{25}, 
Alan~Boss\altaffilmark{26}, 
Edna~Devore\altaffilmark{3}, 
Alan~Gould\altaffilmark{27}, 
Jeffrey~C.~Smith\altaffilmark{3}, 
Robert~L.~Morris\altaffilmark{3}, 
Andrej~Prsa\altaffilmark{28}, 
Timothy~D.~Morton\altaffilmark{21} 
}
\altaffiltext{1}{University of California, Berkeley, CA 94720}
\altaffiltext{2}{NASA Ames Research Center, Moffett Field, CA 94035}
\altaffiltext{3}{SETI Institute/NASA Ames Research Center, Moffett Field, CA 94035}
\altaffiltext{4}{San Jose State University, San Jose, CA 95192}
\altaffiltext{5}{Lowell Observatory, Flagstaff, AZ 86001}
\altaffiltext{6}{Jet Propulsion Laboratory/Caltech, Pasadena, CA 91109}
\altaffiltext{7}{University of Texas, Austin, TX 78712}
\altaffiltext{8}{Harvard-Smithsonian Center for Astrophysics, 60 Garden Street, Cambridge, MA 02138}
\altaffiltext{9}{Las Cumbres Observatory Global Telescope, Goleta, CA 93117}
\altaffiltext{10}{Space Telescope Science Institute, Baltimore, MD 21218}
\altaffiltext{11}{Niels Bohr Institute, Copenhagen University, Denmark}
\altaffiltext{12}{Aarhus University, DK-8000 Aarhus C, Denmark}
\altaffiltext{13}{High Altitude Observatory, National Center for Atmospheric Research, Boulder, CO 80307}
\altaffiltext{14}{NASA Exoplanet Science Institute/Caltech, Pasadena, CA 91125}
\altaffiltext{15}{National Optical Astronomy Observatory, Tucson, AZ 85719}
\altaffiltext{16}{Massachusetts Institute of Technology, Cambridge, MA, 02139}
\altaffiltext{17}{Fermilab Center for Particle Astrophysics, Batavia, IL 60510}
\altaffiltext{18}{University of California, Santa Cruz, CA 95064}
\altaffiltext{19}{Yale University, New Haven, CT 06510}
\altaffiltext{20}{University of Florida, Gainesville, FL 32611}
\altaffiltext{21}{Orbital Sciences Corp., NASA Ames Research Center, Moffett Field, CA 94035}
\altaffiltext{22}{California Institute of Technology, Pasadena, CA 91109}
\altaffiltext{23}{Department of Physics, Broida Hall, University of California, Santa Barbara, CA 93106}
\altaffiltext{24}{Centre for Astrophysics Research, University of Hertfordshire, College Lane, Hatfield, AL10 9AB, England}
\altaffiltext{25}{San Diego State University, San Diego, CA 92182}
\altaffiltext{26}{Carnegie Institution of Washington, Dept.\ of Terrestrial Magnetism, Washington, DC 20015}
\altaffiltext{27}{Lawrence Hall of Science, Berkeley, CA 94720}
\altaffiltext{28}{Villanova University, Dept. of Astronomy and Astrophysics, 800 E Lancaster Ave, Villanova, PA 19085}
\altaffiltext{$\dagger$}{Based in part on observations obtained at the W.~M.~Keck Observatory, which is operated by the University of California and the California Institute of Technology.}

\altaffiltext{*}{To whom correspondence should be addressed.  E-mail: howard@astro.berkeley.edu}

\begin{abstract}
We report the distribution of planets as a function of planet radius, orbital period, 
and stellar effective temperature for orbital periods less than 50 days around Solar-type (GK) stars.   
These results are based on the 1,235 planets (formally ``planet candidates'') from the \Kepler mission 
that include a nearly complete set of detected planets as small as 2 \rearthe.
For each of the 156,000 target stars we assess the detectability of planets as a function of planet radius, \rp, and orbital period, $P$, 
using a measure of the detection efficiency for each star. 
We also correct for the geometric probability of transit, \rstar/$a$.
We consider first  \kepler target stars within the ``solar subset'' having \teff\ = 4100--6100 K, \logg\ = 4.0--4.9,
and Kepler magnitude $\kp < 15$ mag, i.e.\ bright, main sequence GK stars.  
We include only those stars having photometric noise low enough to permit detection of planets down to 2~\rearthe. 
We count planets in small domains of \rp\ and $P$ and divide by the included target stars 
to calculate planet occurrence in each domain.
The resulting occurrence of planets varies by more than three orders of magnitude in the radius-orbital period plane and  
increases substantially down to the smallest radius (2~\rearthe) and out to the longest orbital period (50 days, $\sim$0.25 AU) in our study.  
For $P < 50$ days, the distribution of planet radii is given by a power law, $\mathrm{d}f / \mathrm{d}\log R= k_R R^{\alpha} $ 
with $k_R$ = \kRPL, $\alpha$ = \alphaRPL, and $R$ = \rp/\rearthe.   
This rapid increase in planet occurrence with decreasing planet size agrees with 
the prediction of core-accretion formation, but disagrees with population synthesis models that 
predict a desert at super-Earth and Neptune sizes for close-in orbits. 
Planets with orbital periods shorter than 2 days are extremely rare; 
for $\rp > 2$ \rearth we measure an occurrence of less than $0.001$ planets per star.
For all planets with orbital periods less than 50 days, we measure occurrence of  
\OccPfiftyRtwofour, \OccPfiftyRfoureight, and \OccPfiftyReightthirtytwo\ planets per star 
for planets with radii 2--4, 4--8, and 8--32 \rearthe, in agreement with Doppler surveys.  
We fit occurrence as a function of $P$ to a power law model with an exponential cutoff below a critical period $P_0$.  
For smaller planets, $P_0$ has larger values, suggesting that the ``parking distance'' for migrating planets moves outward 
with decreasing planet size.
We also measured planet occurrence over a broader stellar \teff\ range of 3600--7100~K, spanning M0 to F2 dwarfs. 
Over this range, the occurrence of 2--4~\rearth planets in the \Kepler field linearly increases with decreasing \teff,   
making these small planets seven times more abundant around cool stars (3600--4100 K) than the hottest stars in our sample (6600--7100 K).
\end{abstract}

\section{Introduction}
\label{sec:intro}

The dominant theory for the formation of planets within 20 AU involves the collisions and sticking of planetesimals 
having a rock and ice composition, growing to Earth-size and beyond.  
The presence of gas in the protoplanetary disk allows gravitational accretion of hydrogen, helium and other volatiles, 
with accretion rates depending on gas density and temperature, and hence on location within the disk and its stage of evolution. 
The relevant processes, including inward migration, have been simulated numerically both for individual planet growth and for 
entire populations of planets  \citep{Ida04a,Ida_Lin_2008_V,Mordasini2009, Schlaufman2010, Ida_Lin_2010, Alibert2011}. 

The simulations suggest that most planets form near or beyond the ice line.   
When they reach a critical mass of several Earth-masses (\mearthe) the planets either rapidly spiral inward to the host star 
because of the onset of Type II migration  or undergo runaway gas accretion and become massive gas-giants, 
thus producing a ``planet desert'' \citep{Ida_Lin_2008_IV}.   
The predicted desert resides in the mass range $\sim$1--20 \mearth orbiting inside of $\sim$1 AU, 
with details that vary with assumed behavior of inward planet migration \citep{Ida_Lin_2008_V, Ida_Lin_2010, Alibert2011, Schlaufman2009}.   
Another prediction is that the distribution of planets in the mass/orbital distance plane is fairly uniform for 
masses above the planet desert ($\gtrsim$\,20 \mearthe) and inside of 
$\sim$0.25 AU (periods less than 50 days).  
The majority of the planets in these models reside near or beyond the ice line at $\sim$2 AU 
(well outside of the $P < 50$~days domains analyzed here).  
The mass distribution for these distant planets rises toward super-Earth and Earth-mass 
\citep{Ida_Lin_2008_V,Mordasini09b, Alibert2011}.
These patterns of planet occurrence in the two-parameter space defined by planet masses and orbital periods 
can be directly tested with observations of a statistically large sample of planets orbiting within 1 AU of their host stars.

Two early observational tests of the planet-formation simulations have emerged.  
Using RV-detected planets, \cite{Howard2010b} measured planet occurrence for close-in planets ($P<50$ days) 
with masses that span nearly three orders of magnitude---super-Earths to Jupiters (\msini\ = 3--1000 \mearthe). 
This Eta-Earth Survey focused on 166 G and K dwarfs on the main sequence. 
The survey showed an increasing occurrence, $f$, of planets with decreasing mass, $M$, from 1000 to 3 \mearthe.  
A power law fit to the observed distribution of planet mass gave d\textit{f}/dlog\textit{M} = 0.39\textit{M}$^{-0.48}$. 
Remarkably, the survey revealed a high occurrence of planets in the period range $P$ = 10--50 days and 
mass range \msini = 4--10 \mearthe, precisely within the predicted planet desert.   
Planets with \msini\  = 10--100 \mearth and $P < 20$ days were found to be quite rare.  
Thus, the predicted desert was found to be full of planets and the predicted uniform mass distribution for 
close-in planets above the desert was found to be rising with smaller mass, not flat.   
These discrepancies suggest that current population synthesis models of planet formation around solar-type stars 
are somehow failing to explain the distribution of low-mass planets around solar-type stars.

Accounting for completeness, \cite{Howard2010b} found a planet occurrence  of $15^{+5}_{-4}$\%    
for planets with \msini\ = 3--30 \mearth and $P < 50$ d around main sequence G and K stars.  
In contrast, Mayor et al.\ have asserted a substantially higher planet occurrence of 
30\% $\pm$ 10\% \citep{Mayor09a} or higher with a careful statistical study still in progress.  
Thus, there may be observational discrepancies in planet occurrence  which we expect to be resolved soon.  
Still, there is qualitative agreement between \cite{Howard2010b} and \cite{Mayor09a} 
that the predicted paucity of planets of mass $\sim$1--30 \mearth within 1 AU is not observed, 
as that close-in domain is, in fact, rich with small planets.
The planet candidates from \Keplere, along with a careful assessment of both false positive rates and completeness, 
can add a key independent measure of the occurrence of small planets to compare with the Eta-Earth Survey and Mayor et al.
Formally these objects are ``planet candidates'' as a small percentage will turn out to be false positive detections; 
we often refer to them as ``planets'' below.

The observed occurrence of small planets orbiting close-in matches continuously with the similar analysis by 
\cite{Cumming08} who measured 10.5\% of Solar-type stars hosting a gas-giant planet 
(\msini\ = 100--3000 \mearthe, $P$ = 2--2000 days), for which planet occurrence varies as 
$\mathrm{d}f \propto M^{-0.31\pm0.2} P^{0.26\pm0.1} \, \mathrm{d}\log M \, \mathrm{d}\log P$. 
Thus, the occurrence of giant planets orbiting in 0.5--3 AU seems to attach smoothly to the occurrence of planets 
down to 3 \mearth orbiting within 0.25 AU.   This suggests that the formation and accretion processes are continuous 
in that domain of planet mass and orbital distance, or that the admixture of relevant processes varies continuously 
from 1000 \mearth down to 3 \mearthe.
 
Planet formation theory must also account for remarkable orbital properties of exoplanets.   
The orbital eccentricities span the range $e$ = 0--0.93 and the close-in ``hot Jupiters'' show a wide distribution of alignments 
(or misalignments) with the equatorial plane of the host star \citep[e.g.,][]{Johnson2009, Winn2010, Winn2011, Triaud2010, Morton2010}.    
Thus, standard planet formation theory probably requires additional planet-planet gravitational interactions to 
explain these non-circular and non-coplanar orbits \citep[e.g.][]{Chatterjee_Ford_Rasio_2010, Wu_Lithwick_2011, Nagasawa2008}.

The distribution of planets in the mass/orbital-period plane reveals important clues 
about planet formation and migration.   Here we carry out an analysis of the epochal \Kepler results for 
transiting planet candidates from \cite{Borucki2011} with a careful treatment of the completeness.  
We focus attention on the planets with orbital periods less than 50 days to match the period range that RV surveys are most sensitive to.   
The goals are to measure the occurrence distribution of close-in planets, 
to independently test planet population synthesis models, 
and to check the Doppler RV results of \cite{Howard2010b}.   
While none of the planets or stars are in common between \Kepler and RV surveys, 
we will combine the mass distribution (from RV) and the radius distribution (from \Keplere) to 
constrain the bulk densities of the \textit{types} of planets they have in common.
Planet formation models predict great diversity in the interior structures of planets having Earth-mass to Saturn-mass, 
caused by the various admixtures of rock, water-ice, and H \& He gas.   
Here we attempt to statistically assess planet radii and masses to arrive for the first time 
at the density distribution of planets within 0.25 AU of their host stars.

\section{Selection of Kepler Target Stars and Planet Candidates}
\label{sec:selection}

We seek to determine the occurrence of planets as a function of orbital period, planet radius (from \Keplere) and 
planet mass (from Doppler searches).   
Measuring occurrence using either Doppler or transit techniques suffers from detection efficiency that is a function of 
both the properties of the planet (radius, orbital period) and of the individual stars (notably noise from stellar activity).  
Thus the effective stellar sample from which occurrence may be measured is itself a function of planet properties and 
the quality of the data for each target star.  
A key element of this paper is that {\it only a subset of the target stars are amenable to the detection of planets 
having a certain radius and period.} 

To overcome this challenge posed by planet detection completeness, we construct a two-dimensional space of 
orbital period and planet radius (or mass).  
We divide this space into small domains of specified increments in period and planet radius (or mass) and 
carefully determine the subset of target stars for which the detection of planets in that small domain has high efficiency.  
In that way, each domain of orbital period and planet size (or mass) has its own subsample of target stars that are 
selected {\it a priori}, within which the detected planets can be counted and compared to that number of stars.  
This treatment of detection completeness for each target star was successfully adopted by \cite{Howard2010b} 
in the assessment of planet occurrence as a function of orbital period and planet mass (\msini) from Doppler surveys.  
Here we carry out a similar analysis of occurrence of planets from the \Kepler survey 
in a two-dimensional space of orbital period and planet radius.

\subsection{Winnowing the Kepler Target Stars for High Planet Detectability}

To measure planet occurrence we compare the number of detected planets having some set of properties 
(radii, orbital periods, etc.) to the set of stars from which planets with those properties could have been reliably detected.  
Errors in either the number of planets detected or the number of stars surveyed corrupt the planet occurrence measurement.  
We adopt the philosophy that it is preferable to suffer higher Poisson errors from considering fewer planets and stars than the 
difficult-to-quantify systematic errors caused by studying a larger number of planets and stars with more poorly determined 
detection completeness.  

We begin our winnowing of target stars with the \Kepler Input Catalog \citep[KIC; ][]{Brown2011_kic, KIC2009}.  
In this paper we include only planet candidates found in three data segments (``Quarters'') 
labeled Q0, Q1, and Q2, for which all photometry is published \citep{Borucki2011}. 
Q0 was data commissioning (2--11 May 2009), Q1 includes data from 13 May to 15 June 2009, 
and Q2 includes data from 15 June to 17 September 2009. 
The segments had durations of  9.7, 33.5, and 93 days, respectively.  
\Kepler achieved a duty cycle of greater than 90\%, which almost completely eliminated window function effects \citep{vonBraun2009}.   
A total of 156,097 long cadence targets (30 min integrations) were observed in Q1 and 166,247 targets were observed in Q2, 
with the targets in Q2 being nearly a superset of those in Q1.  In this paper we consider only the ``exoplanet target stars'' 
of which there were 153,196 observed during Q2, and are used for the statistics presented here \citep{Batalha2010_targets}.  
(The remaining \Kepler targets in Q2 were evolved stars, not suitable for sensitive planet detection.)   
The few percent changes in the planet-search target stars are not significant here as Q2 data dominate the planet detectability.  
The KIC contains stellar \teff\ and radii (\rstar) that are based on four visible-light magnitudes ($g, r, i, z$) and a fifth, 
\textit{D51}, calibrated with model atmospheres, and \textit{JHK} IR magnitudes \citep{Brown2011_kic}.

The photometric calibrations yield \teff\ reliable to $\pm$135~K (rms) and surface gravity \logg\ reliable to $\pm$0.25~dex (rms), 
based on a comparison of KIC values to results of high resolution spectra obtained with the Keck I telescope and 
LTE analysis \citep{Brown2011_kic}.   
Stellar radii are estimated from \teff\ and \logg\ and carry an uncertainty of 0.13 dex, i.e.\ 35\% rms \citep{Brown2011_kic}.  
There is a concern that values of \logg\ for subgiants are systematically overestimated, 
leading to stellar radii that are smaller than their true radii perhaps by as much as a factor of two.  
One should be concerned that a magnitude-limited survey such as \Kepler may favor slightly evolved stars, 
implying systematic underestimates of stellar radii, an effect worth considering at the interpretation stage of this work.  
The quoted planet radii may be too small by as much as a factor of two for evolved stars.   
We adopt these KIC values for stellar \teff\ and \rstar\ from the KIC and their associated uncertainties, following \cite{Borucki2011}.  
The stellar metallicities are poorly known.  
The KIC is available on the Multi-Mission Archive at the Space Telescope Science Institute (MAST) 
website\footnote{http://archive.stsci.edu/kepler/}.

\begin{figure}
\includegraphics[width=0.48\textwidth]{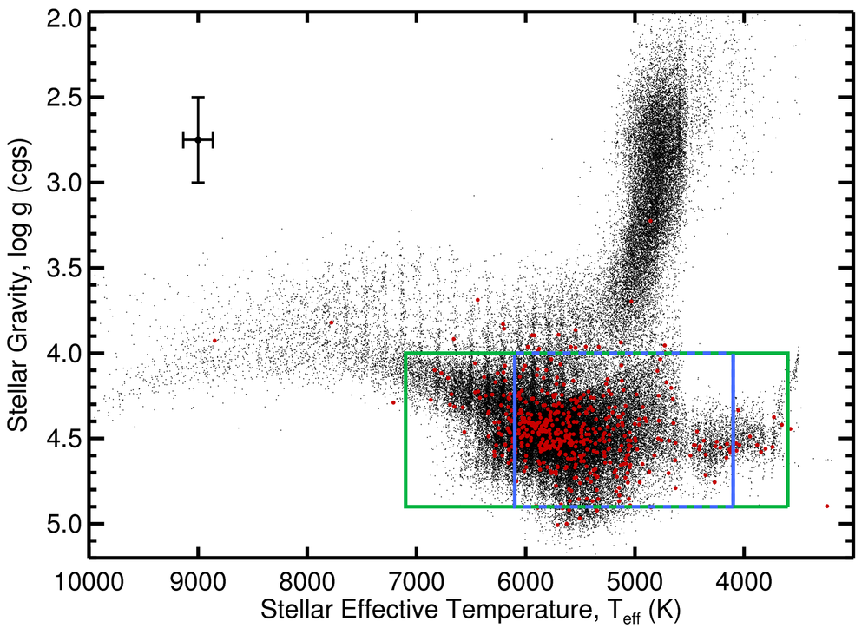}
\caption{\kepler target stars (small black dots) and \kepler stars with planet candidates (red dots) 
plotted as a function of \teff\ and \logg\ from the KIC.  
Only bright stars ($\kp < 15$) are shown and considered in this study.  
The inner blue rectangle marks the ``solar subset'' (\teff\ = 4100--6100 K and \logg\ = 4.0--4.9) 
of main sequence G and K stars considered for most of this study.
This domain contains \Nst\ stars with \Npl\ planet candidates. 
In Section \ref{sec:teff} we consider planet occurrence as a function of \teff.  
For that analysis we consider a broader range of \teff\ = 3600--7100 K (green outer rectangle).
The error bars in the upper left show the typical uncertainties of 135 K in \teff\ and 0.25 dex in \logg.}
\label{fig:logg_teff}
\end{figure}

In this paper, we primarily consider \Kepler target stars having properties in the core of the \Kepler mission, 
namely bright solar-type main sequence stars.  Specifically, we consider only Kepler target stars within this 
domain of the H-R diagram: \teff\ = 4100--6100 K, \logg\ = 4.0--4.9, 
and Kepler magnitude $\kp < 15$ mag (Table \ref{tab:sample_properties}).  
These parameters select for the brightest half of the GK-type target stars (the other half being fainter, $\kp > 15$ mag), 
as shown in Figure \ref{fig:logg_teff}.   
The goal is to limit our study to main sequence GK stars well characterized in the KIC \citep{Brown2011_kic} 
and to provide a stellar sample that is a close match to that of \cite{Howard2010b}, 
offering an opportunity for a comparison of the radii and masses from the two surveys.   
The brightness limit of \kp\ $<$ 15 promotes high photometric signal-to-noise ratios, needed to detect the smaller planets.   
These three criteria in \teff, \logg, and \kp\ seem, at first glance, to be quite modest, representing the core target stars 
in the \Kepler mission.  \textit{Yet these three stellar criteria yield a subsample of only \Nst\ target stars, 
roughly one third of the total Kepler sample.}   
In this study, we consider only this subset of \Kepler stars and the associated planet candidates detected among them.  

\begin{deluxetable}{lc}
\tabletypesize{\footnotesize}
\tablecaption{Properties of Stellar and Planetary Samples
\label{tab:sample_properties}}
\tablewidth{0pt}
\tablehead{
  \colhead{Parameter}   & 
  \colhead{Value} 
}
\startdata
Stellar effective temperature, \teff  & 4100--6100 K \\
Stellar gravity, \logg\ (cgs) &  4.0--4.9 \\
Kepler magnitude, $\kp$ & $<$ 15 \\
Number of stars, $n_\star$ & \Nst \\
Orbital period, $P$ & $<$ 50 days\\
Planet radius, \rp & 2--32 \rearth \\
Detection threshold, SNR (90 days) & $>$ 10 \\
Number of planet candidates, $n_\mathrm{pl}$ & \Npl \\
\enddata
\end{deluxetable}

\subsection{Winnowing Kepler Target Stars by Detectable Planet Radius and Period}

We further restrict the \Kepler stellar sample by including only those stars with high enough photometric quality 
to permit detection of planets of a specified radius and orbital period.   
To begin, we consider differential domains in the two-dimensional space of planet radius and orbital period.  
For each differential domain, only a subset of the \Kepler target stars have sufficient photometric quality to permit detection of such a planet.   
In effect, the survey for such specific planets is carried out {\it only} among those stars having photometric quality 
so high that the transit signals stand out easily (literally by eye).   
For photometric quality we adopt the metric of the signal-to-noise ratio (SNR) of the transit signal integrated over a 
90 day photometric time series.  
We define SNR to be the transit depth divided by the uncertainty in that depth due to photometric noise (to be defined quantitatively below).

We set a threshold, SNR $>$ 10, which is higher than that (SNR $>$ 7.0) adopted by \cite{Borucki2011}, 
lending our study an even higher standard of detection.  
Thus, we restrict our sample of stars so strongly that planets of a specified radius and orbital period are rarely, 
if ever, missed by the  ``Transiting Planet Search'' \citep[TPS; ][]{Jenkins2010_TPS} pipeline.   
Moreover, we base our SNR criterion on just a single 90 day quarter of \Kepler photometry.  
This conservatively demands that the photometric pipeline detect transits only during a single pointing of the telescope.   
(The CCD pixels that a particular star falls on change quarterly as \Kepler is rolled by 90 degrees to maintain solar illumination.)
As noted in \cite{Borucki2011}, the photometric pipeline does not yet have the capability to stitch together multiple quarters 
of photometry and search for transits.  
In contrast, the SNR quoted in \cite{Borucki2011} was based on the totality of photometry, Q0--Q5 (approximately one year in duration).  
Thus we are setting a threshold that is considerably more stringent than in \cite{Borucki2011}, 
i.e.\ including target stars of the quietest photometric behavior.   
The goal, described in more detail below, is to establish a subset of \Kepler target stars for which the 
detection efficiency of planets (of specified radius and orbital period) is close to 100\%.   

Finally, we restrict our study to orbital periods under 50 days.  
All criteria by which \Kepler stars are retained in our study are given in Table \ref{tab:sample_properties}.  
As demonstrated below, these restrictions on SNR $>$ 10 and on orbital period ($P< 50$ days) yield a final subsample 
of \Kepler targets for which very few planet candidates will be missed by the current \Kepler photometric pipeline as the 
transit signals both overwhelm the noise and repeat multiple times (for $P<50$ days).

We explored the adoption of two measures of photometric SNR for each \Kepler star, one taken directly from 
\cite{Borucki2011} and the other using the so-called Combined Differential Photometric Precision ($\sigma_{\rm CDPP}$) 
which is the empirical RMS noise in bins of a specified time interval, 
coming from the \Kepler pipeline.  Actually, \cite{Borucki2011} derived their SNR values from $\sigma_{\rm CDPP}$, 
integrated over all transits in Q0--Q5.  
We employed the measured $\sigma_{\rm CDPP}$ for time intervals of 3~hr and compared the resulting SNR from \citet{Borucki2011} 
for transits to those we computed from the basic $\sigma_{\rm CDPP}$.   
These values agreed well (understandably, accounting for the use of a total SNR from all five quarters in \citet{Borucki2011}).  
Thus, we adopted the basic 3 hr $\sigma_{\rm CDPP}$ for each target star as the origin of our noise measure.

Each Kepler target star has its own measured RMS noise level, $\sigma_{\rm CDPP}$.  
Typical 3~hr $\sigma_{\rm CDPP}$ values are 30--300 parts per million (ppm), 
as shown in Figure 1 of \cite{Jenkins_lc_Data}, albeit for 6~hr time bins.   
Clearly, the photometrically noisiest target stars are less amenable to the detection of small planets, which we treat below. 
The noise has three sources.  
One is simply Poisson errors from the finite number of photons received, dependent on the star's brightness, 
causing fainter stars to have higher $\sigma_{\rm CDPP}$.  
This photon-limited photometric noise is represented by the lower envelope of the noise as a function of magnitude in 
Figure 1 of \cite{Jenkins_lc_Data}.  
A second noise source stems from stellar surface physics including spots, convective overshoot and turbulence (granulation), 
acoustic p-modes, and magnetic effects arising from plage regions and reconnection events.   
A third noise source stems from excess image motion in Q0, Q1, and Q2 stemming from use of variable guide stars that
have now been dropped.
In Q2 the presence of bulk drift corrected by four re-pointings of the bore sight, plus a safe
mode followed by an unusually large thermal recovery also contributed.  
The measured $\sigma_{\rm CDPP}$ accounts for all such sources, as well as any unmentioned since it is an empirical measure.

Using $\sigma_{\rm CDPP}$ for each target star, we define SNR integrated over all transits as,
\begin{equation}
\mathrm{SNR} = \frac{\delta}{\sigma_\mathrm{CDPP}}\sqrt{\frac{n_{\mathrm{tr}} \cdot t_\mathrm{dur}}{3\,\mathrm{hr}}}.
\label{eq:SNR}
\end{equation}
Here $\delta = \rp^2/\rstar^2$ is the photometric depth of a central transit of a 
planet of radius \rp\ transiting a star of radius \rstar, $n_{\mathrm{tr}}$ is the number of 
transits observed in a 90 day quarter, $t_\mathrm{dur}$ is the transit duration, 
and the factor of 3 hr accounts for the duration over which $\sigma_{\rm CDPP}$ was measured.  
We include only those stars yielding SNR $>$ 10, for a given specified transit depth and orbital period.
The threshold imposes such a stringent selection of target stars that few planets are missed by the 
\Kepler Transiting Planet Search (TPS)  pipeline.  
Our planet occurrence analysis below assumes that (nearly) all planets with $\rp > 2 $\rearth that 
meet the above SNR criteria have been detected by the \Kepler pipeline and are included in \cite{Borucki2011}.  
The \Kepler team is currently engaged in a considerable study of the completeness of 
the \Kepler pipeline by injecting simulated transit signals into pipeline at the CCD pixel level and 
measuring the recovery rate of those signals as a function of SNR and other parameters.  
In advance of the results of this major numerical experiment, 
we demonstrate detection completeness of SNR $>$ 10 signals in two ways.

\begin{figure*}
\includegraphics[width=1.0\textwidth]{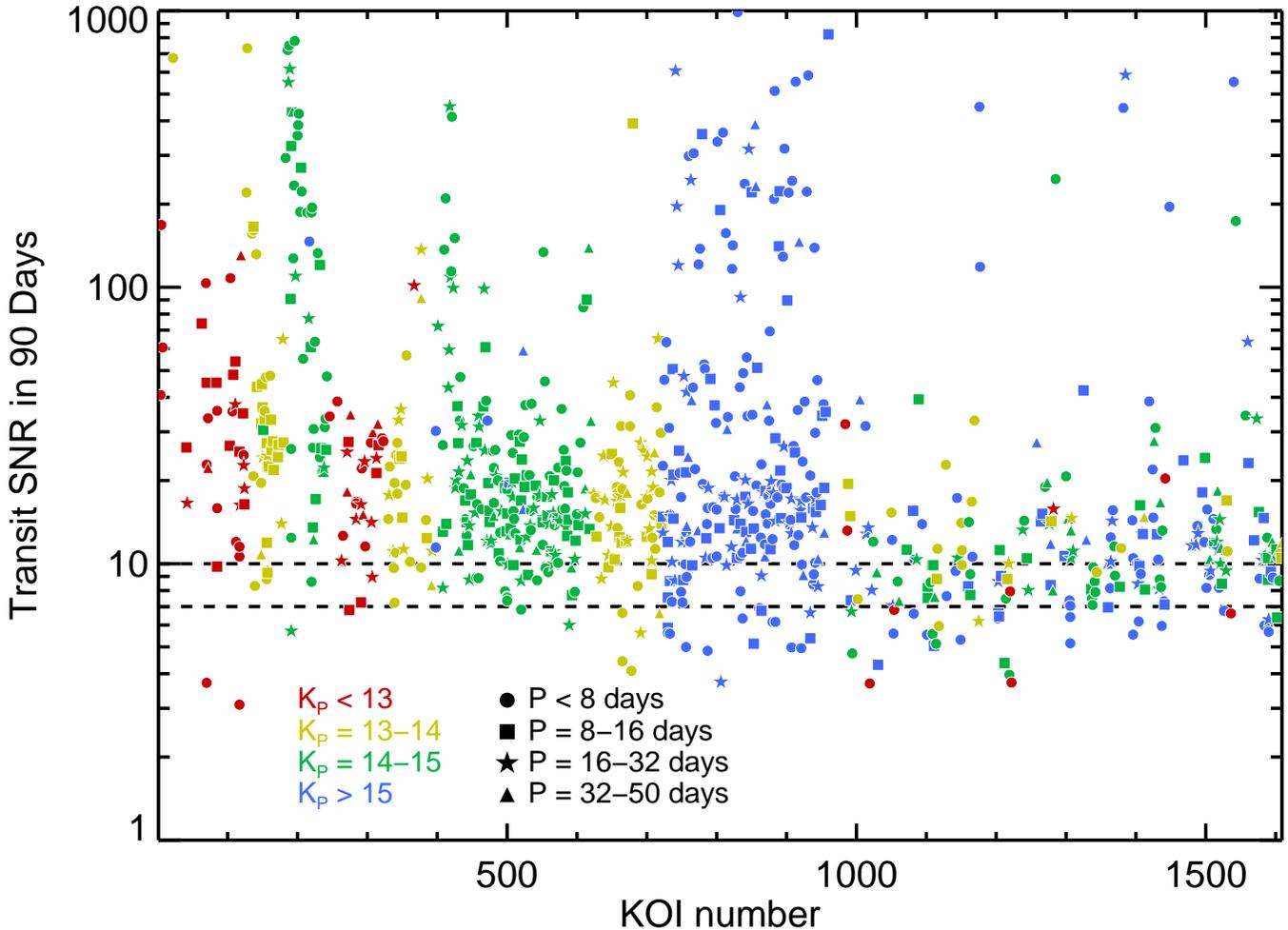}
\caption{Signal to Noise Ratio (SNR) of detected transits 
during a 90 day quarter.  See text for definition of SNR.  
Only planets orbiting stars with \teff\ = 4100--6100 K and \logg\ = 4.0--4.9 are shown.  
Planets orbiting stars with Kepler magnitudes $\kp < $13, 13--14, 14--15, and $> 15$ 
are shown in red, yellow, green, and blue, respectively.  
Planets with orbital periods $P < 8$, 8--16, 16--32, and $> 32$ days are 
shown as filled circles, squares, five-point stars, and triangles, respectively.
Our analysis considers only transits with SNR $>$ 10 (upper dashed line).}
\label{fig:koi_num_nsigma}
\end{figure*}

First, Figure \ref{fig:koi_num_nsigma} shows the SNR of detected transits as a function of \Kepler Object of Interest (KOI) number.   
The \Kepler photometry and TPS pipeline detects planet candidates over the course of months as data arrive.  
There is a learning curve involved with this process, as both software matures and human intervention is tuned \citep{Rowe2010}.  
As a result, the obvious (high SNR) planet candidates are issued low KOI numbers as they are detected early in the mission.  
The shallower transits, relative to noise, are identified later as they require more data, and are issued larger KOI numbers.  
Thus KOI number is a rough proxy for the time required to accumulate enough photometry to identify the planet candidate.  
Among the KOIs 1050--1600, much less vetting was done, 
and indeed we rejected five planet candidates (KOIs 1187, 1227, 1387, 1391, and 1465)
reported in \citet{Borucki2011} based on 
both V-shaped light curves and at least one other property indicating a likely eclipsing binary.

Figure \ref{fig:koi_num_nsigma} shows that the early KOIs, 1--1000, had a wide range of SNR values spanning 7--1000, 
as the first transit signals had a variety of depths.  
KOIs 400--1000 correspond to pipeline detections of transit planet candidates around target stars as faint as 15th mag and fainter. 
The more recent transit identifications of KOIs 1000--1600 exhibit far fewer transits with SNR $>$ 20 
and about half of these new KOIs have SNR $<$ 10, below our threshold for inclusion.  
Apparently most newly identified KOIs have SNR $<$ 20, and few planets remain to be found with $P<50$ days and SNR $>$ 20.  
\textit{Figure \ref{fig:koi_num_nsigma} suggests that the great majority of planet candidates with $P< 50$ days and SNR $>$ 10 
have already been identified by the \Kepler pipeline.}   
This apparent asymptotic success in the detection of SNR $>$ 10 transits is enabled by our orbital period limit of 50 days 
which is considerably less than the duration of a quarter (90 days).   
The current \Kepler pipeline for identifying transits within a single 90 day quarter is more robust than the multi-quarter transit search.  
For such short periods, at least two transits typically occur within one quarter. 
Moreover, when such planet candidates appear during another quarter, the short period planets are quickly confirmed.   
We suspect that for periods greater than 90 days, many more planet candidates are yet to be identified by \Keplere.   
Thus, this study restricts itself to $P < 50$ days in part because of the demonstrated completeness of detection for such short periods.

\begin{figure*}
\includegraphics[width=1.0\textwidth]{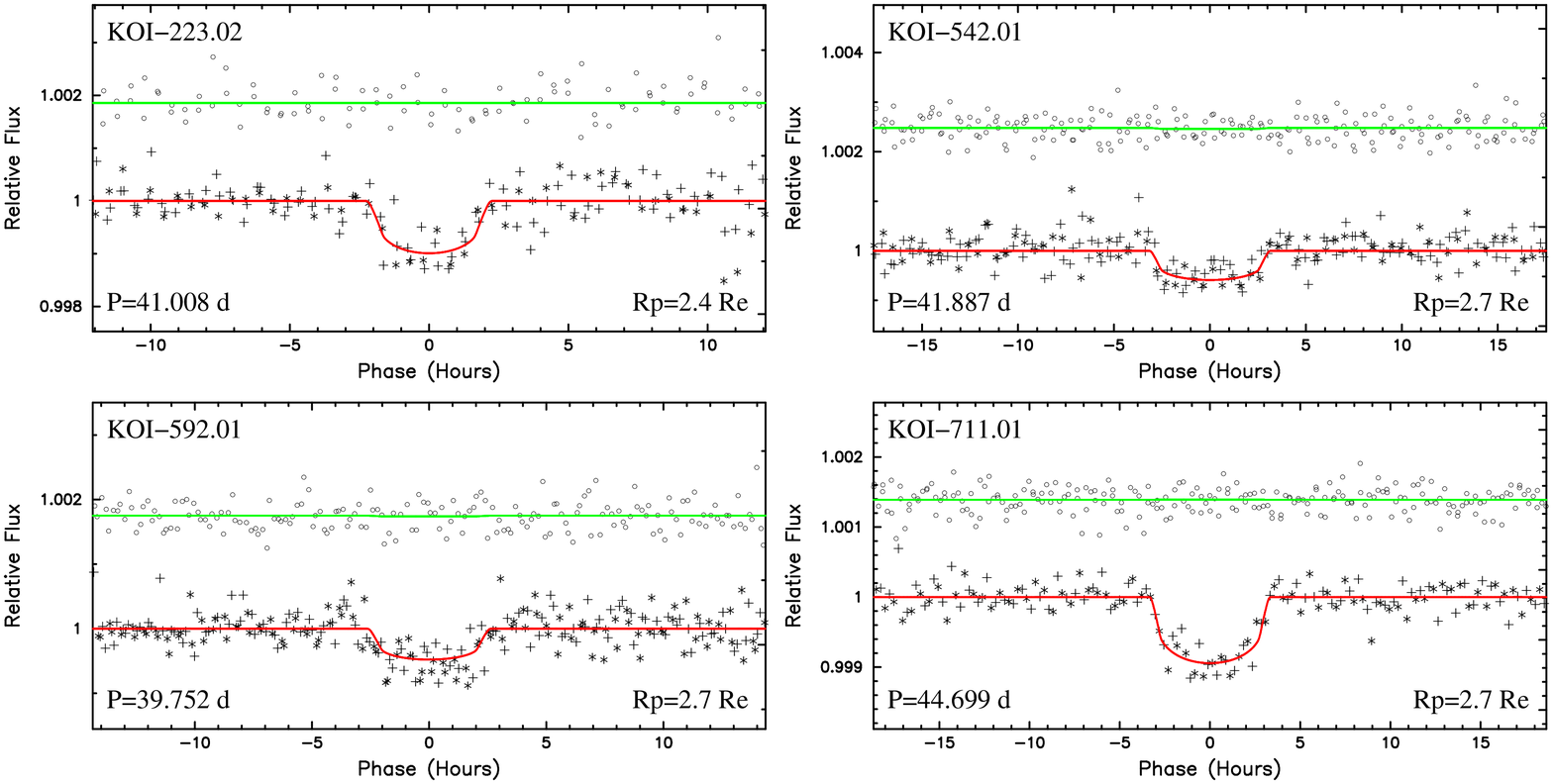}
\caption{Four representative light curves of planet candidates with \rp\ = 2--3 \rearth and 
$P$ = 30--50 days, the domain of most challenging detection in this study.  
See Table \ref{tab:light_curves} for planetary and stellar properties.  
In each panel, the transit light curve (lower, red trace) and photometric measurements 180 degrees out of phase 
(upper, green trace) are shown with the best-fit model overlaid.  
Plus and star symbols show alternating transits. 
Only photometry from Q2 is displayed.  
These light curves demonstrate the data quality for the 
lowest SNR planet candidates included in this study, 
most with a transit depth of only $\sim$10 times greater than the uncertainty in the mean depth due to noise.  
Still, the transits are clearly visible to the eye.  
The \Kepler pipeline is unlikely to miss many of these planet candidates, despite their being in the least detectable 
domain of the study.   
This indicates the security of these detections and the high completeness of such planet candidates, 
in support of Figure \ref{fig:koi_num_nsigma}.}
\label{fig:light_curves}
\end{figure*}

We examined the light curves themselves for a second demonstration of nearly complete detection efficiency 
of planet candidates with $P < 50$ days, $\rp > 2$ \rearthe, and SNR $>$ 10.
Figure \ref{fig:light_curves} shows four representative light curves of planet candidates whose properties are 
listed in Table \ref{tab:light_curves}.
All four have small radii of 2--3 \rearth and ``long'' periods of 30--50 days, 
the most difficult domain for planet detection in this study (the lower right corner of Figure \ref{fig:rp_per}, discussed below).  
The SNR values are near the threshold value of $\sim$10; in fact, one planet candidate (KOI 592.01) 
has a SNR of 9.7 and is therefore conservatively excluded from this study.  
\textit{The four light curves show how clearly such transits stand out, indicating the high detection completeness 
of planets down to 2 \rearth and $P<50$ days for the SNR $>$ 10 threshold we adopted.}   

\begin{deluxetable}{rcccccc}
\tabletypesize{\footnotesize}
\tablecaption{Properties of Planet Candidates in Figure \ref{fig:light_curves}
\label{tab:light_curves}}
\tablewidth{0pt}
\tablehead{
  \colhead{KOI}   & 
  \colhead{\kp} &
  \colhead{$R_\star$} &
  \colhead{\rp} &
  \colhead{$P$} &
  \colhead{SNR} &
  \colhead{SNR}  \\
  \colhead{}   & 
  \colhead{(mag)} &
  \colhead{(\rsune)} &
  \colhead{(\rearthe)} &
  \colhead{(days)} &
  \colhead{(Q0--Q5)} &
  \colhead{(90 days)} 
}
\startdata
223.02 & 14.7 & 0.74 & 2.40 & 41.0 & 25 & 12.3 \\
542.01 & 14.4 & 1.13 & 2.70 & 41.9 & 21 & 11.2 \\
592.01 & 14.3 & 1.08 & 2.70 & 39.8 & 19 & \phn9.7 \\
711.01 & 14.0 & 1.00 & 2.74 & 44.7 & 34 & 25.3 
\enddata
\end{deluxetable}

\subsection{Identifying Kepler Planet Candidates}

We adopt the \Kepler planet candidates and their orbital periods and planet radii from Table 2 of \cite{Borucki2011}, 
with two exceptions.  
First, we exclude the five KOIs noted above that are likely to be false positives.   
Second, we exclude KOIs that orbit ``unclassified'' KIC stars 
(identified with ``\teff\ Flag'' = 1 in Table 1 of  \cite{Borucki2011}).  
We measure planet occurrence only around stars with well defined stellar parameters from the KIC.  

To summarize the \cite{Borucki2011} results, photometry at roughly 100 ppm levels in 29.4 minute integrations allows detection of repeated, 
brief drops in stellar brightness caused by planet transits across the star.  The technical specifics of the instrument, photometry, 
and transit detection are described in \cite{Borucki10b, Koch2010_mission, Jenkins_lc_Data, Jenkins2010_TPS, Caldwell2010_inst}.
We begin the identification of planet candidates based on those revealed in public \Kepler photometric data (Q0--Q2).
This data release contains 997 stars with a total of 1,235 planetary candidates that show transit-like signatures, 
all with some follow-up work that could not rule out the planet hypothesis \citep{Gautier2010_followup}.
\cite{Borucki2011} includes three planets discovered in the \Kepler field before launch: 
TrES-2b \citep{ODonovan2006}, 
HAT-P-7b \citep{Pal2008}, 
and HAT-P-11b \citep{Bakos09}.  
We are including only those planet candidates that meet two SNR standards:  
They must have SNR $>$ 10 in one quarter alone and they must have SNR $>$ 7 in all quarters.  
The former standard should guarantee the latter, but this double-standard reinforces the quality of the planet candidates.   

As this data release contains 136 days of photometric data, with only a few small windows of down time, 
most planet candidates with periods under 50 days have exhibited two or more transits.  
The multiple transits for $P<$ 50 days offer relatively secure candidates, periods, and radii, provided by the repeated transit light curves.  
For P $<$ 40 days, \Kepler has detected typically three or more transits in the publicly available data.  
Moreover, in \cite{Borucki2011} the periods, radii, and ephemerides are based on the full set of \Kepler data obtained in Q0--5, 
constituting over one year of photometric data.   
Thus, planet candidates with periods under 50 days are securely detected with multiple transits.  
They have improved SNR in the light curves from the full set of data available to the \Kepler team, 
offering excellent verification, radii, and periods for short period planets.  

\subsection{False Positives}

We expect that some of the planet candidates reported in \cite{Borucki2011} are actually false positives.   
These would be mostly background eclipsing binaries diluted by the foreground star.  
They may also be background stars orbited by a transiting planet of larger radius, 
but diluted by the light of the foreground star mimicking a smaller planet.   
False positives can also occur from gravitationally bound companion stars that are eclipsing binaries or have larger transiting planets.   
We expect that false positive probabilities will be estimated for most planet candidates in \citet{Borucki2011} 
using ``BLENDER'' \citep{Torres_blender2011}. 

In the mean time, the false positive rate has been estimated carefully by \cite{Morton2011}.   
They find the false positive probability for candidates that pass the standard vetting gates to be less than 10\% and normally closer to 5\%.  
In particular, the \Kepler vetting process included a difference analysis between CCD images taken in and out of transit, 
allowing direct detection of the pixel that contains the eclipsing binary, if any.   
This vetting process found that $\sim$12\% of the original planet candidates were indeed eclipsing binaries in neighboring pixels, 
and these were deemed false positives and removed from Table 2 of \cite{Borucki2011}.   
This process leaves only the one pixel itself, with a half-width of 2 arcsec within which any eclipsing binary must reside.  
As 12\% of the planet candidates had an eclipsing binary within the $\sim$10 pixels total of the photometric aperture, 
the rate of eclipsing binaries hidden behind the remaining one pixel is likely to be $\sim$1.2\%, a small probability of false positives.   
The bound, hierarchical eclipsing binaries were estimated by \cite{Morton2011}, finding another few percent may be such false positives, 
yielding a total false positive probability of $\sim$5--10\%.  
\cite{Morton2011} note that the false positive probability depends on transit depth $\delta$, galactic latitude $b$, and \kp.  
Using their ``detailed framework'' and computing the false positive probability for each of the \Npl\ planet candidates 
among our ``solar subset'' (Table \ref{tab:sample_properties}), we estimate that 22 planet candidates are actually false positives.\footnote{We 
note that while the precise details of these estimates depend on \textit{a priori} assumptions of the overall planet occurrence rate 
(which we conservatively take to be 20\%) and of the planet radius distribution (which follows Figure 5 of \cite{Morton2011}), 
the overall low false positive probability is controlled by the relative scarcity of blend scenarios compared to planets.  
We also note that these estimates do not account for uncertainties in \rstar, which may result in some jovian-sized candidates actually 
being M dwarfs eclipsing subgiant stars.}  
The resulting false positive rate of 5\% is on the low end of the 5--10\% estimate above because we restricted our stellar sample 
to bright main sequence stars and planet sample to $\rp > 2$ \rearthe.
We do not expect this low false positive rate to substantially impact our statistical results below.

Nearly all of the KOIs reported in \cite{Borucki2011} are formally ``planet candidates'', 
absent planet validation \citep{Torres_blender2011} or mass determination 
\citep{Borucki2010_Kepler-4, Koch2010_Kepler-5, Dunham2010_Kepler-6, Latham2010_Kepler-7, Jenkins_Kepler-8, 
Holman2010_Kepler-9, Batalha2011, Lissauer2011}. 
For simplicity we will refer to all KOIs as ``planets'', bearing in mind that a small percentage will turn out to be false positives.

\section{Planet Occurrence}
\label{sec:occurrence}

\begin{figure*}
\includegraphics[width=1.0\textwidth]{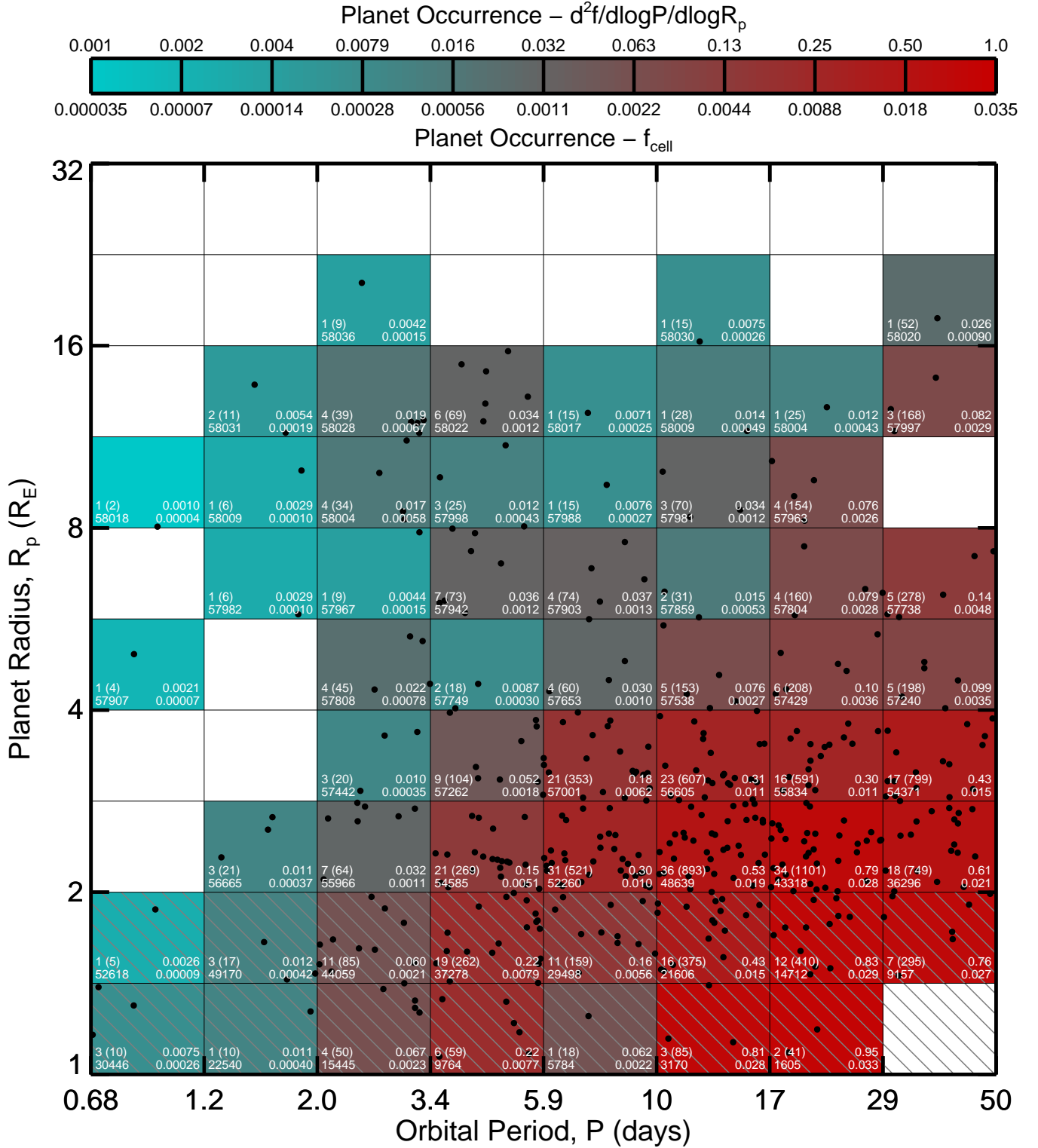}
\caption{Planet occurrence as a function of planet radius and orbital period for $P < 50$ days.  
Planet occurrence spans more than three orders of magnitude 
and increases substantially for longer orbital periods and smaller planet radii.
Planets detected by \Kepler having SNR $>$ 10 are shown as black dots.  
The phase space is divided into a grid of  logarithmically spaced cells within which planet occurrence is computed.  
Only stars in the ``solar subset'' (see selection criteria in Table \ref{tab:sample_properties}) were used to compute occurrence.  
Cell color indicates planet occurrence with the color scale on the top in two sets of units, 
occurrence per cell and occurrence per logarithmic area unit.  
White cells contain no detected planets.
Planet occurrence measurements are incomplete and likely contain systematic errors in the hatched region ($\rp <  2$ \rearthe).
Annotations in white text within each cell list occurrence statistics: 
\textit{upper left}---the number of detected planets with SNR $>$ 10, $n_{\mathrm{pl,cell}}$, 
and in parentheses the number of augmented planets correcting for non-transiting geometries, $n_{\mathrm{pl,aug,cell}}$;
\textit{lower left}---the number of stars surveyed by \Kepler around which a hypothetical transiting planet with \rp\ and $P$ values 
from the middle of the cell could be detected with SNR $>$ 10; 
\textit{lower right}---$f_{\rm cell}$, planet occurrence, corrected for geometry and detection incompleteness; 
\textit{upper right}---$d^2f/\mathrm{d} \log_{10}P / \mathrm{d} \log_{10}R_{\rm p}$, 
planet occurrence per logarithmic area unit ($\mathrm{d} \log_{10}P\ \mathrm{d} \log_{10}R_{\rm p}$ = \BinsLogArea\ grid cells).}
\label{fig:rp_per}
\end{figure*}

We define planet occurrence, $f$, as the fraction of a defined population of stars (in \teff, \logg, $\kp$) 
having planets within a domain of planet radius and period, including all orbital inclinations.
We computed planet occurrence as a function of planet radius and orbital period 
in the grid of cells in Figure \ref{fig:rp_per}.
Within each cell we counted the number of planets detected by \Kepler for 
the subset of stars surveyed with sufficient precision to compute the local planet occurrence, $f_{\rm cell}$.  
Our treatment corrects for planets not detected by \kepler because of non-transiting orbital inclinations 
and because of insufficient photometric precision.   

The average planet occurrence within a confined cell of \rp\ and $P$ is 
\begin{equation}
f_{\mathrm{cell}} = \sum_{j=1}^{n_{\mathrm{pl,cell}}} \frac{1/p_j}{n_{\star,j}},
\label{eq:eta}
\end{equation}
where the sum is over all detected planets within the cell that have SNR $>$ 10.  
In the numerator, $p_j = (R_{\star}/a)_j$ is the \textit{a priori} probability of a transiting orientation of planet $j$.  
Each individual planet is augmented in its contribution to the planet count by a factor of $a/R_{\star}$ 
to account for the number of planets 
with similar radii and periods that are not detected because of non-transiting geometries.
For each planet, its specific value of $(a/R_{\star})_j$ is used, not the average $a/R_{\star}$ of the cell in which it resides.
Each scaled semi-major axis $(a/R_{\star})_j$ is measured directly from \kepler photometry 
and is not the ratio of two quantities, $a_j$ and $R_{\star,j}$, separately measured with lower precision.  
In the denominator, $n_{\star,j}$ is the number of stars whose physical properties and photometric stability 
are sufficient so that a planet of radius $R_{{\rm p},j}$ and period $P_j$ 
would have been detected with SNR $>$ 10 as defined by equation (\ref{eq:SNR}).
Note that our requirement for SNR $>$ 10 is applied to the numerator (the planets that count toward the occurrence rate) 
and the denominator (the stars around which those planets could have been detected) of equation (\ref{eq:eta}).  

While Figure \ref{fig:rp_per} does not show error estimates for $f_{\mathrm{cell}}$, 
we compute them with binomial statistics and use them in the analysis that follows.  
We calculate the binomial probability distribution of drawing $n_{\mathrm{pl,cell}}$ 
planets from $n_{\mathrm{\star,eff,cell}} = n_{\mathrm{pl,cell}}/f_{\mathrm{cell}}$ ``effective'' stars.  
The $\pm1\sigma$ errors in $f_{\mathrm{cell}}$ are computed from the 
15.9 and 84.1 percentile levels in the cumulative binomial distribution.
Note that $n_{\mathrm{pl,cell}}$ is typically a small number (in Figure \ref{fig:rp_per}, $n_{\mathrm{pl,cell}}$ 
has a range of 1--36 detected planets) so the errors within individual cells can be significant.  
These errors and the corresponding occurrence  fluctuations between adjacent cells
average out when cells are binned together to compute occurrence  as a function of radius or period.  
Also note that our error estimates only account for random errors and not systematic effects.

Figure \ref{fig:rp_per} contains numerical annotations to help digest the wealth of planet occurrence information.  
In the lower left of each cell is $n_{\mathrm{\star,mid-cell}}$, 
the number of \Kepler targets with sufficient $\sigma_{\rm CDPP}$ such 
that a central transit of a planet with \rp\ and $P$ values from the middle of the cell could have been detected with SNR $>$ 10.
Above this, we list $n_{\mathrm{pl,cell}}$ followed by $n_{\mathrm{pl,aug,cell}}$ in parentheses.  
$n_{\mathrm{pl,aug,cell}}$ is the total extrapolated number of planets in each cell after correcting 
for the \textit{a priori} transit probability for each planet,
\begin{equation}
n_{\mathrm{pl,aug,cell}} =  \sum_{j=1}^{n_{\mathrm{pl}}}1/p_j.
\end{equation}
The annotation in the lower right of each cell is $f_{\mathrm{cell}}$.
The reader can quickly check that planet occurrence is computed 
correctly by verifying that $f_{\mathrm{cell}} \approx n_{\mathrm{pl,aug,cell}}/ n_{\mathrm{\star,mid-cell}}$; 
planet occurrence is the ratio of the number of planets to the number of stars searched.\footnote{This 
approximate expression for $f_{\mathrm{cell}}$ breaks down in cells where the number of stars 
with SNR $>$ 10 ($n_{\star}$) varies rapidly across the cell.    
Equation (\ref{eq:eta}) computes planet occurrence locally using $n_{\star}$ for the specific radius and period of each detected planet, 
while the $n_{\mathrm{\star,mid-cell}}$ listed in the annotations applies to \rp\ and $P$ in the middle of the cell.  
Thus, planet occurrence is more poorly determined in regions of Figure \ref{fig:rp_per} 
where the detection completeness varies rapidly and/or the detected planets are clustered in one section of the cell.  
These poorly measured regions typically have $\rp < 2$~\rearth and longer orbital periods.}
Finally, the annotation in the top right of each cell is $f_{\mathrm{cell}}$ in units of occurrence per 
$\mathrm{d} \log_{10}P\ \mathrm{d} \log_{10}R_p$ (occurrence per factor of ten in \rp\ and $P$), 
a unit that is independent of the choice of cell size.  
There are \BinsLogArea\ grid cells per unit of $\mathrm{d} \log_{10}P\ \mathrm{d} \log_{10}R_p$; 
that is, a region whose edges span factors of ten in \rp\ and $P$ has \BinsLogArea\ grid cells of the size shown in Figure \ref{fig:rp_per}.
Each cell spans a factor of $\sqrt{2}$ in \rp\ and a factor of $5^{1/3}$ in $P$.

The distribution of planet occurrence in Figure \ref{fig:rp_per} offers remarkable clues about the processes of 
planet formation, migration, and evolution.
\textit{Planet occurrence increases substantially with decreasing planet radius and increasing orbital period.}  
Planets larger than 1.5 times the size of Jupiter ($\rp > 16$ \rearthe) are extremely rare.  
Planets with $P \lesssim 2$ days are similarly rare.
Because of incompleteness, we tread with caution for planets with $\rp = $ 1--2 \rearthe, 
but note that these planets have a occurrence similar to 
planets with \rp\ = 2--4 \rearthe.  
Their actual occurrence could be higher due to incompleteness of the pipeline at identifying the smallest planets  
or lower due to a higher rate of false positives.

Planet multiplicity complicates our measurements of planet occurrence.  
We interpret $f_{\mathrm{cell}}$ as the fraction of stars having a planet in the narrow range of $P$ and \rp\ that define a particular cell.  
With few exceptions, stars are not orbited by planets with nearly the same radii and periods.  
However, when we apply equation (\ref{eq:eta}) to larger domains of the radius-period plane, 
for example by marginalizing over $P$ (Section \ref{sec:occ_rp}) or over \rp\ (Section \ref{sec:occ_per}), 
the same star can be counted multiple times in equation (\ref{eq:eta}) if multiple planets fall within that larger domain of \rp\ and $P$.  
Thus, our occurrence measurements are actually of the mean number of planets per star 
meeting some criteria, rather than than fraction of stars having at least one planet that meet  those criteria.  
When the rate of planet multiplicity within a domain is low, these two quantities are nearly equal.

\begin{deluxetable}{cccc}
\tabletypesize{\footnotesize}
\tablecaption{Planet Multiplicity vs. Planet Size
\label{tab:multi}}
\tablewidth{0pt}
\tablehead{
\colhead{} &
\multicolumn{3}{c}{Fraction of planet hosts with a second planet \ldots} \\
\cline{2-4} \\
\vspace{-0.2in}\\
\colhead{\rp\ (\rearthe)} & 
\colhead{in same \rp\ range} &
\colhead{within $\frac{1}{2}$\rp--2\rp} &
\colhead{with any \rp}
}
\startdata
1.0--1.4 & 0.05 & 0.16 & 0.26 \\
1.4--2.0 & 0.09 & 0.25 & 0.27 \\
2.0--2.8 & 0.08 & 0.23 & 0.25 \\
2.8--4.0 & 0.12 & 0.28 & 0.30 \\
4.0--5.6 & 0.04 & 0.09 & 0.13 \\
5.6--8.0 & 0.04 & 0.09 & 0.13 \\
\phn8.0--11.3 & 0.00 & 0.06 & 0.06 \\
11.3--16.0 & 0.00 & 0.00 & 0.06\\
16.0--22.6 & 0.00 & 0.00 & 0.00
\enddata
\end{deluxetable}

The \Npl\ planets in our solar subset of stars (Table \ref{tab:sample_properties}) orbit a total of \NplUniq\ stars.  
The fraction of planets in multi-transiting systems is 0.27 and the fraction of host stars with multiple 
transiting planets is 0.15. 
In Table \ref{tab:multi} we list three measures of planet multiplicity for the planetary systems within the 
solar subset (Table \ref{tab:sample_properties}).  For each of the \rp\ ranges in Figure \ref{fig:rp_per} 
we list the fraction of hosts stars with more than one planet in the specified \rp\ range, 
the fraction of hosts with one planet in the \rp\ range and a second planet with a radius 
within a factor of two of the first planet's,
and the fraction with one planet in the \rp\ range and a second planet having any \rp.  

Table \ref{tab:multi} suggests that multiplicity is common.   
\citet{Lissauer2011b} noted that the multi-planet systems observed by Kepler have relatively low mutual inclinations 
(typically a few degrees) suggesting a significant correlation of inclinations. 
Converting our measurements of the mean number of planets per star to the fraction of stars having at least one planet 
requires an understanding of the underlying multiplicity and inclination distributions.
Such an analysis is attempted by \citet{Lissauer2011b}, but is beyond the scope of this paper. 

It is worth identifying additional sources of error and simplifying assumptions in our methods.  
The largest source of error stems directly from 35\% rms uncertainty in \rstar\ from the KIC, 
which propagates directly to 35\% uncertainty in \rp.  
We assumed a central transit over the full stellar diameter in equation (\ref{eq:eta}).  
For randomly distributed transiting orientations, the average duration is reduced to $\pi/4$ times the duration of a central transit. 
Thus, this correction reduces our SNR in equation (\ref{eq:SNR}) by a factor of $\sqrt{\pi/4}$, 
i.e.\ a true signal-to-noise ratio threshold of 8.8 instead of 10.0.  This is still a very conservative detection threshold.  
Additionally, our method does not account for the small fraction of transits that are grazing and have reduced significance.  
We assumed perfect $\sqrt{t}$ scaling for $\sigma_{\rm CDPP}$ values computed for 3 hr intervals.
This may underestimate $\sigma_{\rm CDPP}$ for a 6 hr interval (approximately the duration of a $P$ = 50 day transit) 
by $\sim$10\%.
These are minor corrections and affect the numerator and denominator of equation (\ref{eq:eta}) nearly equally.  

\subsection{Occurrence as a Function of Planet Radius}
\label{sec:occ_rp}

\begin{figure}
\includegraphics[width=0.48\textwidth]{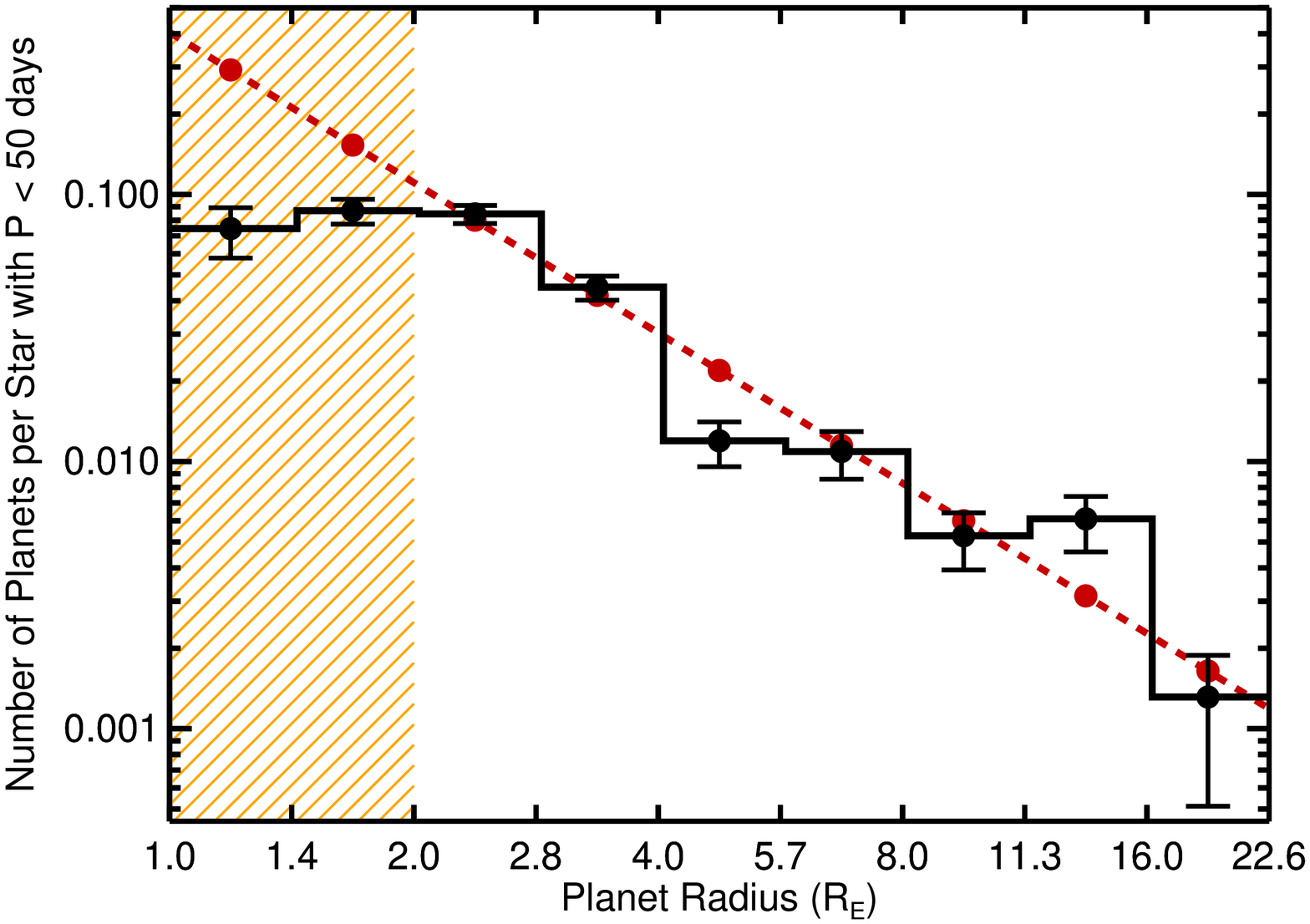}
\includegraphics[width=0.48\textwidth]{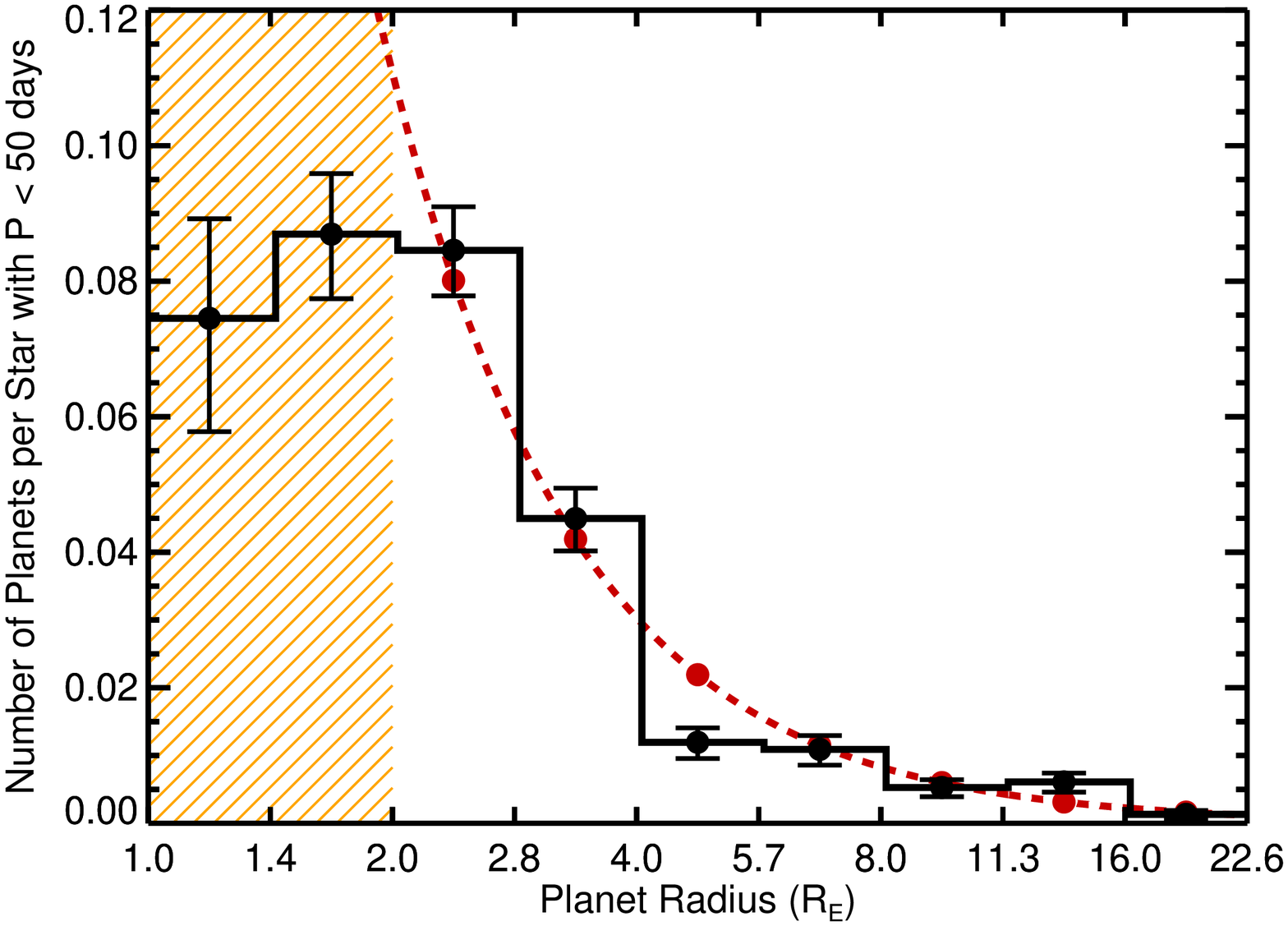}
\caption{Planet occurrence as a function of planet radius for planets with $P < 50$ days (black filled circles and histogram).  
The top and bottom panels show the same planet occurrence measurements on logarithmic and linear scales.
Only GK stars consistent with the selection criteria in Table \ref{tab:sample_properties} were used to compute occurrence.  
These measurements are the sum of occurrence values along rows in Figure \ref{fig:rp_per}.
Estimates of planet occurrence are incomplete in the hatched region ($\rp < 2$ \rearthe).  
Error bars indicate statistical uncertainties and do not include systematic effects, 
which are particularly important for $\rp < 2$ \rearthe.
No planets with radii of 22.6--32 \rearth were detected (see top row of cells in Figure \ref{fig:rp_per}).
A power law fit to occurrence measurements for \rp\ = 2--22.6 \rearth (red filled circles and dashed line) 
demonstrates that close-in planet occurrence increases substantially with decreasing planet radius.}
\label{fig:rp_dist}
\end{figure}

Planet occurrence varies by three orders of magnitude in the radius-period plane (Figure \ref{fig:rp_per}).  
To isolate the dependence on these parameters, we first considered planet occurrence as a function of planet radius, 
marginalizing over all planets with $P < 50$ days.  
We computed occurrence using equation (\ref{eq:eta}) for cells with the ranges of radii in Figure \ref{fig:rp_per} 
but for all periods less than 50 days.  
This is equivalent to summing the occurrence values in Figure \ref{fig:rp_per} along rows of cells 
to obtain the occurrence for all planets in a radius interval with $P < 50$ days.  
The resulting distribution of planet radii (Figure \ref{fig:rp_dist}) increases substantially with decreasing \rp.

We modeled this distribution of planet occurrence with planet radius as a power law of the form
\begin{equation}
\frac{\mathrm{d}f(R)}{\mathrm{d}\log R} = k_R R^{\alpha}.
\label{eq:occ_rp}
\end{equation}
Here $\mathrm{d}f(R)/\mathrm{d}\log R$ is the mean number of planets having $P < 50$ days per star 
in a $\log_{10}$ radius interval centered on $R$ (in \rearthe),
$k_R$ is a normalization constant, and $\alpha$  is the power law exponent.  
To estimate these parameters, we used measurements from the 2--22.7 \rearth bins 
because of incompleteness at smaller radii and a lack of planets at larger radii.  
We fit equation (\ref{eq:occ_rp}) using a maximum likelihood method \citep{Johnson2010}. 
Each radius interval contains an estimate of the planet fraction, 
$F_i = \mathrm{d}f(R_i)/\mathrm{d}\log R$, based on a number
of planet detections made from among an effective number of target stars, such
that the probability of $F_i$ is given by the binomial distribution
\begin{equation}
p(F_i|n_{\rm pl}, n_{\rm nd}) = F_i^{n_{\rm pl}} (1-F_i)^{n_{\rm nd}}
\end{equation}
where $n_{\rm pl}$ is the number of planets detected in a specified radius interval (marginalized over period,   
$n_{\rm nd} \equiv n_{\rm pl}/f_{\rm cell} - n_{\rm pl}$ is the effective number of non-detections per radius interval, 
and $f_{\rm cell}$ is the estimate of planet occurrence over the marginalized radius interval obtained 
from equation (\ref{eq:eta}). 
The planet fraction varies as a function of the mean planet radius $R_{{\rm p},i}$ in each bin, and the
best-fitting parameters can be obtained by maximizing the
probability of all  bins using the model in equation (\ref{eq:occ_rp}):
\begin{equation}
\mathcal{L} = \prod_{i=1}^{n_{\rm bin}} p(F(R_{{\rm p},i})).
\end{equation}
In practice the likelihood becomes vanishingly small away
from the best-fitting parameters, so we evaluate the logarithm of the likelihood 
\begin{eqnarray}
\ln{\mathcal{L}} &=& \sum_{i=1}^{n_{\rm bin}} \ln{p(F(R_{{\rm p},i}))} \\ \nonumber
                            &=& \sum_{i=1}^{n_{\rm bin}} n_{{\rm pl}, i} (\ln{k_R} + \alpha \ln{R_{{\rm p},i}}) + n_{{\rm nd}, i} \ln{(1-k_R R_{{\rm p},i}^\alpha)}. \\ \nonumber
\end{eqnarray}
We calculate $\ln{\mathcal{L}}$ over a uniform grid in $k_R$ and
$\alpha$. 
The resulting posterior probability distribution is strongly covariant in $\alpha$ and $k_R$.  
Marginalizing over each parameter, we find $\alpha$ = \alphaRPL\ and $k_R$ = \kRPL, 
where the best-fit values are the median of the marginalized 1-dimensional parameter distributions 
and the error bars are the 15.9 and 84.1 percentile levels.

\citet{Howard2010b} found a power law planet \textit{mass} function, 
d\textit{f}/dlog\textit{M} = $k^{\prime}$\textit{M}$^{\alpha^{\prime}}$, 
with $k^{\prime} = 0.39^{+0.27}_{-0.16}$ and  $\alpha^{\prime} = -0.48^{+0.12}_{-0.14}$
for periods $P < 50$ days and masses \msini\ = 3--1000 \mearthe.
We explore planet densities and the mapping of \rp\ to \msini\ in Section \ref{sec:density}.

\subsection{Occurrence as a Function of Orbital Period}
\label{sec:occ_per}

\begin{figure}
\includegraphics[width=0.48\textwidth]{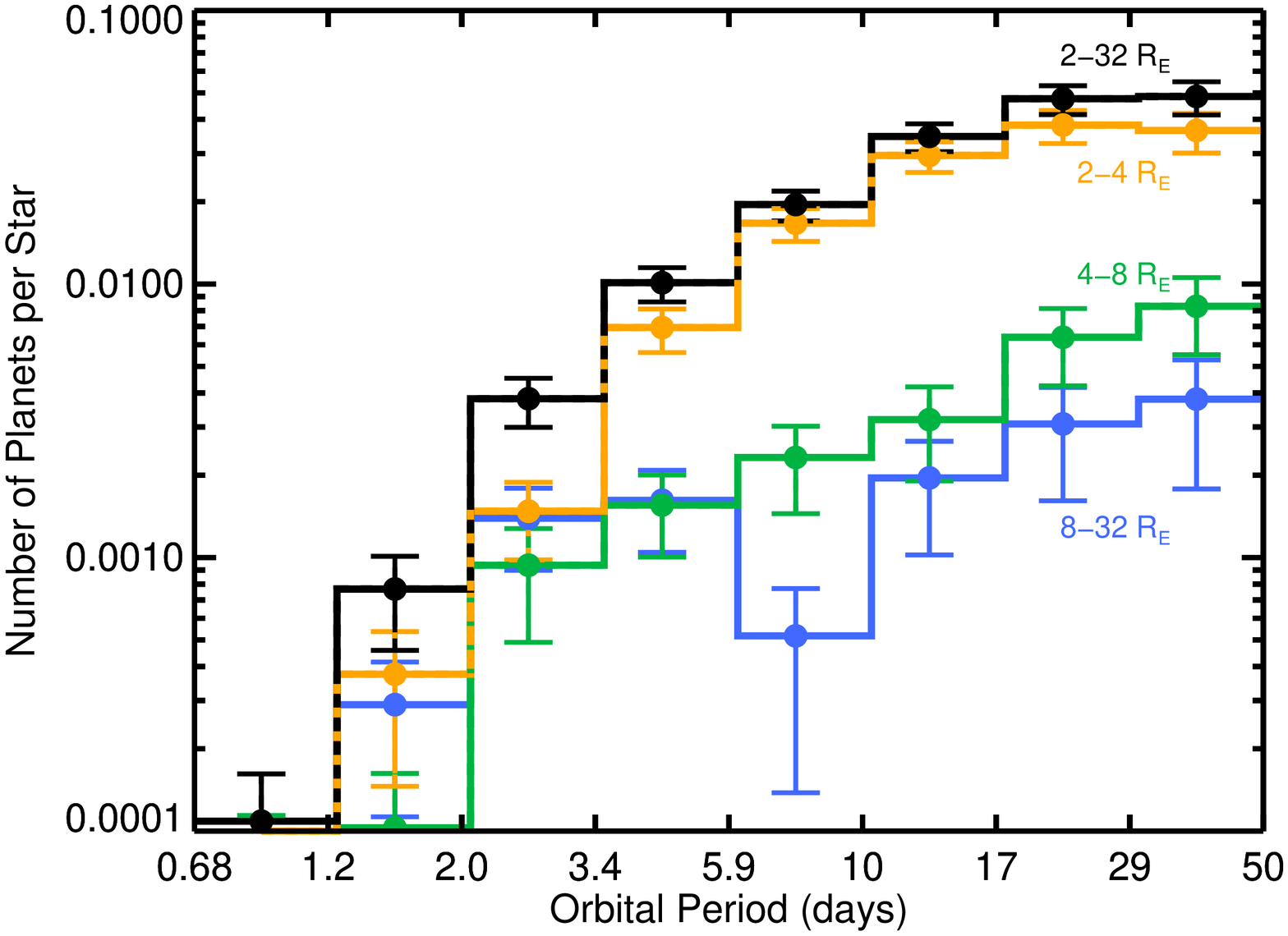}
\includegraphics[width=0.48\textwidth]{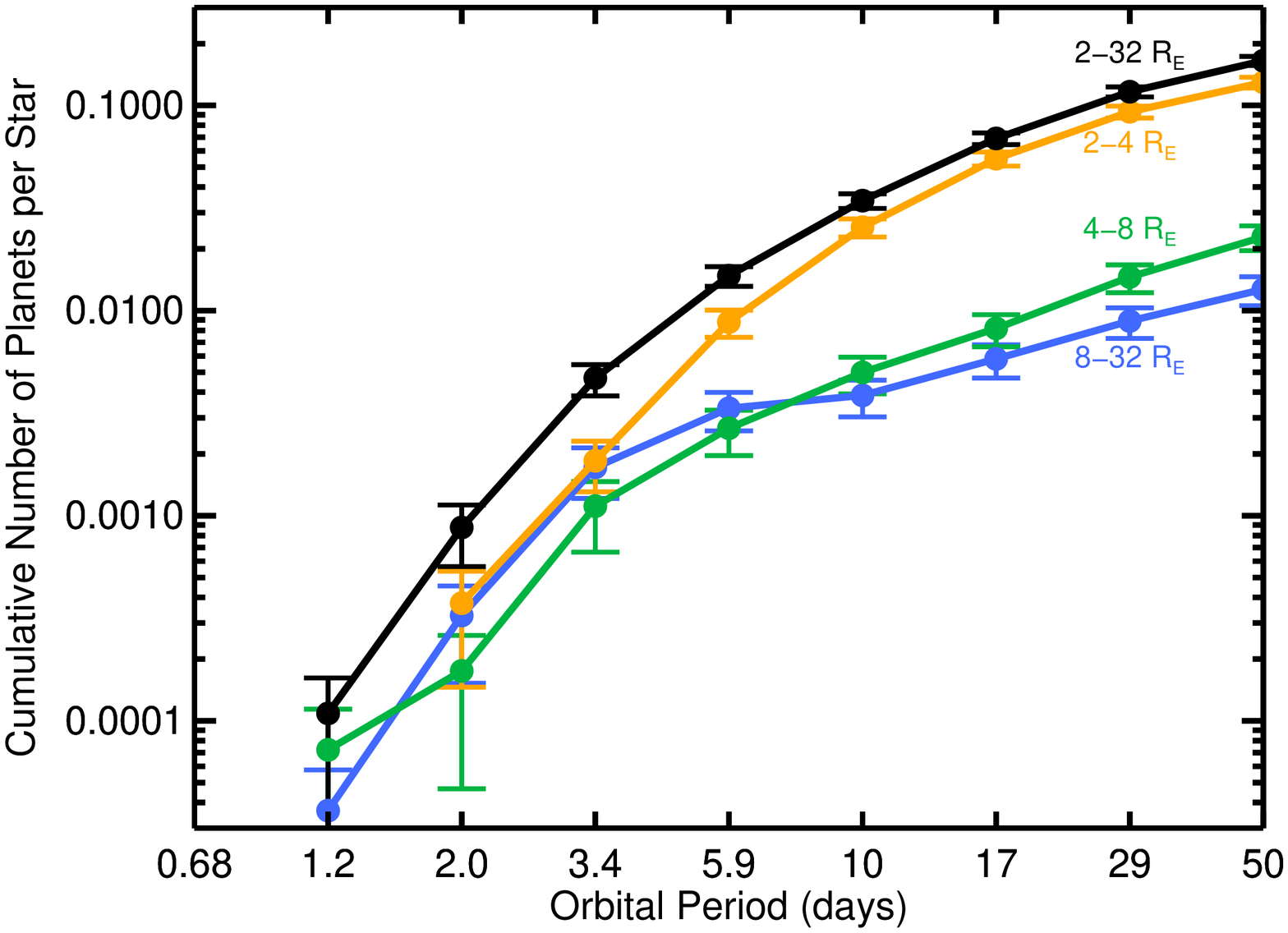}
\caption{Planet occurrence (top panel) and cumulative planet occurrence  (bottom panel) as a function of orbital period.  
The occurrence of planets with radii of 2--32 \rearth (black), 2--4 \rearth (orange), 4-8 \rearth (green), 8--32 \rearth (blue) 
are each depicted.  
Only stars consistent with the selection criteria in Table \ref{tab:sample_properties} were used to compute occurrence.  
Occurrence for planets with $\rp < $ 2~\rearth\ is not shown due to incompleteness. 
The lower panel (cumulative planet occurrence) is the sum of occurrence values in the top panel out to the specified period.}
\label{fig:per_dist}
\end{figure}

We computed planet occurrence as a function of orbital period using equation (\ref{eq:eta}).  
We considered this period dependence for ranges of planet radii ($\rp = 2$--4, 4--8, and 8--32 \rearthe).  
This is equivalent to summing the occurrence values in Figure \ref{fig:rp_per} along two adjacent columns of cells 
to obtain the occurrence for all planets in specified radius ranges.  
Figure \ref{fig:per_dist} shows that planet occurrence increases substantially with increasing orbital period, 
particularly for the smallest planets with \rp\ = 2--4 \rearth.  

For $P <  2$ days, planets of all radii in our study ($>$2 \rearthe) are extremely rare 
with an occurrence of $< 0.001$ planets per star.  
Extending to slightly longer orbital periods, 
hot Jupiters ($P < 10$ days, $\rp = 8$--32 \rearthe) are also rare in the \kepler survey.  
We measure an occurrence of only \OccPtenReightthirtytwo\ planets per star, as listed in Table \ref{tab:rp_per}.  
That occurrence value is based on \kp\ $<$ 15 and the other restrictions that define of the ``solar subset'' (Table \ref{tab:sample_properties}).    
Expanding our stellar sample out to $\kp < 16$, but keeping the other selection criteria constant, 
we find a hot Jupiter occurrence of \OccPtenReightthirtytwoMag\ planets per star.  
This fraction is more robust as it is less sensitive to Poisson errors and our concern about detection incompleteness for \kp\ $>$ 15 
vanishes for hot Jupiters that typically produce SNR $>$ 1000 signals.
\citet{Marcy_Japan_05} found an occurrence of $0.012 \pm 0.001$\ for hot Jupiters 
($a < 0.1$ AU, $P \lesssim 12 d$) around FGK dwarfs in the Solar neighborhood (within 50 pc).  
\textit{Thus, the occurrence of hot Jupiters in the \Kepler field is only 40\% that in the Solar neighborhood.} 
One might worry that our definition of $\rp > 8$ \rearth excludes some hot Jupiters detected by 
RV surveys.  For \kp\ $<$ 16 and the same \teff\ and \logg\ criteria,  
we find an occurrence of \OccPtenRfivethirtytwoMag, which is still 40\% lower than the RV measurement.

However,  we do see modest evidence among the \Kepler giant planets of the pile-up of hot Jupiters at orbital periods 
near 3 days (Figures~\ref{fig:rp_per} and \ref{fig:per_dist}) 
as is dramatically obvious from Doppler surveys of stars in the Solar neighborhood \citep{Marcy08,Wright2009}.
These massive, close-in planets are detected with high completeness in both Doppler and \kepler techniques 
(including the geometrical factor for \Keplere), so the different occurrence values are real.
We are unable to explain this difference, although a paucity of metal-rich stars in the \Kepler sample is one possible explanation. 
Unfortunately, the metallicities of \Kepler stars from KIC photometry are inadequate to test this hypothesis \citep{Brown2011_kic}.  
A future spectroscopic study of \Kepler stars with LTE analysis similar to \citet{Valenti05} offers a possible test.  
In additional to the metallicity difference, the stellar populations may have different \teff\ distributions, 
despite having similar \teff\ ranges.
\citet{Johnson2010} found that giant planet occurrence correlates with both stellar metallicity and stellar mass (for which \teff\ is a proxy).
A full study of the occurrence of hot Jupiters is beyond the scope of this paper, but we note that 
other photometric surveys for transiting hot Jupiters orbiting stars outside of the stellar neighborhood 
have measured reduced planets occurrence \citep{Gilliland2000, Weldrake2008, Gould2006}.

\begin{deluxetable}{lcc}
\tabletypesize{\footnotesize}
\tablecaption{Planet Occurrence for GK Dwarfs
\label{tab:rp_per}}
\tablewidth{0pt}
\tablehead{
  \colhead{\rp (\rearthe)} &
  \colhead{$P$ $<$ 10 days} &
  \colhead{$P$ $<$ 50 days} 
}
\startdata
\vspace*{0.05in}
2--4 \rearth & \OccPtenRtwofour & \OccPfiftyRtwofour \\
\vspace*{0.05in}
4--8 \rearth & \OccPtenRfoureight & \OccPfiftyRfoureight \\
\vspace*{0.05in}
8--32 \rearth & \OccPtenReightthirtytwo & \OccPfiftyReightthirtytwo \\
\vspace*{0.05in}
2--32 \rearth & \OccPtenRtwothirtytwo & \OccPfiftyRtwothirtytwo 
\enddata
\end{deluxetable}

The occurrence  of smaller planets with radii \rp\ = 2--4 \rearth 
rises substantially with increasing $P$ out to $\sim$10 days 
and then rises slowly or plateaus when viewed in a log-log plot 
(orange histogram, top panel of Figure \ref{fig:per_dist}).    
Out to 50 days we estimate an occurrence of \OccPfiftyRtwofour\ planets per star.  
Small planets in this radius range account for  approximately three quarters of the planets 
in our study, corrected for incompleteness.

The occurrence distributions in the top panel of Figure \ref{fig:per_dist} have shapes that are 
more complicated than simple power laws.  Occurrence falls off rapidly at short periods.  
We fit each of these distributions to a power law with an exponential cutoff, 
\begin{equation}
\frac{\mathrm{d}f(P)}{\mathrm{d}\log P} = k_P P^{\beta} \left(1-e^{-(P/P_0)^\gamma}\right). 
\label{eq:powerlaw_cutoff}
\end{equation}
This function behaves like a power law with exponent $\beta$ and normalization $k_P$ for $P \gg P_0$.  
For periods $P$ (in days) near and below the cutoff period $P_0$, $f(P)$ falls off exponentially.  
The sharpness of this transition is governed by $\gamma$.  
Thus the parameters of equation (\ref{eq:powerlaw_cutoff}) 
measure the slope of the power law planet occurrence distribution for 
``longer'' orbital periods as well as the transition period and sharpness of that transition.

We fit equation (\ref{eq:powerlaw_cutoff}) to the four ranges of radii shown in Figure \ref{fig:per_dist} (top panel) 
and list the best-fit parameters in Table \ref{tab:rp_per_fit}.  
We note that $\beta > 0$ for all planet radii considered, i.e.\ planet occurrence increases with $\log P$.  
For the largest planets (\rp\ = 8--32 \rearthe), $\beta$ = \betaPPLReightthirtytwo\ is consistent with the power law 
occurrence distribution derived by \citet{Cumming08} for gas giant planets with periods of 
2--2000 days, $\mathrm{d}f \propto M^{-0.31\pm0.2} P^{0.26\pm0.1} \, \mathrm{d}\log M \, \mathrm{d}\log P$

\begin{figure}
\includegraphics[width=0.48\textwidth]{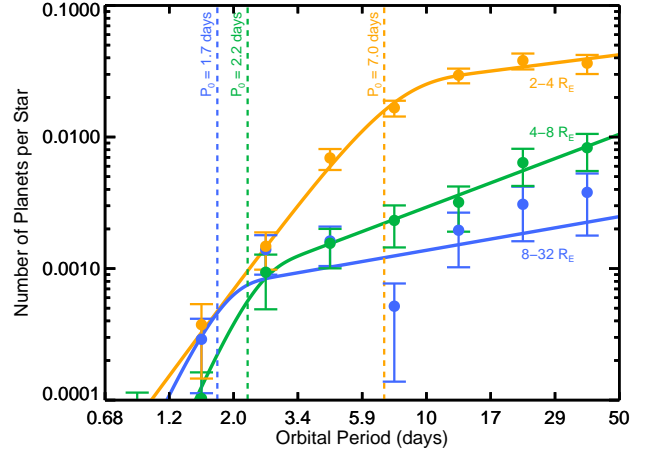}
\caption{Measured planet occurrence (filled circles) as a function of orbital period with best-fit models (solid curves) overlaid.  
These models are power laws with exponential cutoffs below a characteristic period, 
$P_0$ (see text and equation \ref{eq:powerlaw_cutoff}).  
$P_0$ increases with decreasing planet radius, 
suggesting that the migration and parking mechanism that deposits planets close-in depends on planet radius.  
Colors correspond to the same ranges of radii as in Figure \ref{fig:per_dist}.
The occurrence measurements (filled circles) are the same as in Figure \ref{fig:per_dist}, 
however for clarity the 2--32 \rearth measurements and fit are excluded here.
As before, only stars in the solar subset (Table \ref{tab:sample_properties}) and planets with $\rp > $ 2 \rearth 
were used to compute occurrence.}
\label{fig:per_dist2}
\end{figure}

\begin{deluxetable}{ccccc}
\tabletypesize{\footnotesize}
\tablecaption{Best-fit Parameters of Cutoff Power Law Model
\label{tab:rp_per_fit}}
\tablewidth{0pt}
\tablehead{
  \colhead{\rp} &
  \colhead{$k_P$} &
  \colhead{$\beta$} &
  \colhead{$P_0$} &
  \colhead{$\gamma$} \\
  \colhead{(\rearthe)} &
  \colhead{} &
  \colhead{} &
  \colhead{(days)} &
  \colhead{} 
}
\startdata
\vspace*{0.05in}
2--4 \rearthe\phn & \kPPLRtwofour & \betaPPLRtwofour & \PzeroPPLRtwofour & \gammaPPLRtwofour \\
\vspace*{0.05in}
4--8 \rearthe\phn & \kPPLRfoureight & \betaPPLRfoureight & \PzeroPPLRfoureight& \gammaPPLRfoureight \\
\vspace*{0.05in}
8--32 \rearthe & \kPPLReightthirtytwo& \betaPPLReightthirtytwo& \PzeroPPLReightthirtytwo & \gammaPPLReightthirtytwo \\
\vspace*{0.05in}
2--32 \rearthe & \kPPLRtwothirtytwo& \betaPPLRtwothirtytwo & \PzeroPPLRtwothirtytwo & \gammaPPLRtwothirtytwo
\enddata
\end{deluxetable}

$P_0$ and $\gamma$ can be interpreted as  tracers of the migration and stopping mechanisms that 
deposited planets at the closest orbital distances.  
With decreasing planet radius, $P_0$ increases and $\gamma$ decreases, shifting the cutoff period outward 
and making the transition less sharp.
Thus, gas giant planets (\rp\ = 8--32 \rearthe) on average migrate closer to their host stars ($P_0$ is small)
and the stopping mechanism is abrupt ($\gamma$ is large).  
On the other hand, the smallest planets in our study have a distribution of orbital distances (and periods) 
with a characteristic stopping distance farther out and a less abrupt fall-off close-in.  

The normalization constant $k_P$ is highly correlated with the other parameters of equation (\ref{eq:powerlaw_cutoff}).  
A more robust normalization is provided by the requirement that the integrated occurrence to $P = 50$ days 
is given in Table~\ref{tab:rp_per}.


\section{Stellar Effective Temperature}
\label{sec:teff}
\subsection{Planet Occurrence}

In the previous section we considered only GK stars with properties consistent with those listed 
in Table \ref{tab:sample_properties}.
In particular, only stars with \teff\ = 4100--6100 K were used to compute planet occurrence.  
Here we expand this range to 3600--7100 K and measure occurrence as a function of \teff.
This expanded set includes stars as cool as M0 and as hot as F2.  
For \teff\ outside of this range there are too few stars to compute occurrence with reasonable errors.  
We use the same cuts on brightness ($\kp < 15$) and gravity (\logg\ = 4.0--4.9) as before.  
We also used the photometric noise $\sigma_{\rm CDPP}$ values (as before) to compute the 
fraction of target stars around which each detected planet could have been detected with SNR $\geq$ 10.
This ensures that planet detectability down to sizes of 2 \rearth will be close to 100\%, 
for all of these included target stars independent of their \teff.

We computed planet occurrence using the same techniques as in the previous section, namely equation (\ref{eq:eta}).  
We subdivided the stars and their associated planets into 500 K bins of \teff. 
We further subdivided the sample by planet radius, considering different ranges of \rp\ 
(2--4, 4--8, 8--32, and 2--32 \rearthe) separately.  
In summary, we computed planet occurrence as a function of \teff\ for several ranges of \rp\  
and in all cases we considered all planets with $P < 50$ days.

\begin{figure}
\includegraphics[width=0.48\textwidth]{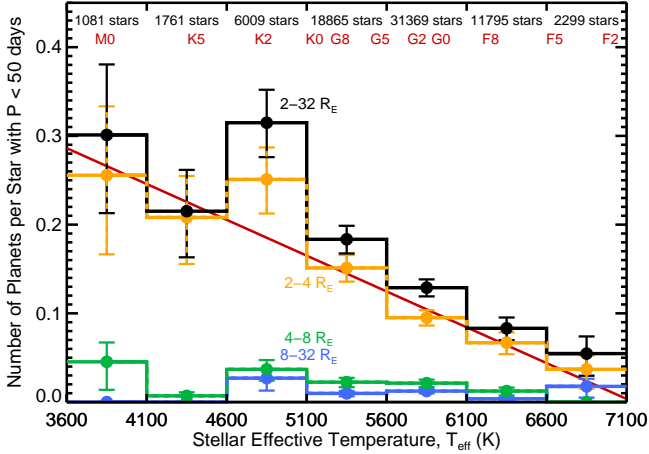}
\caption{Planet occurrence as a function of stellar effective temperature \teff. 
Histogram colors refer to planets with the same ranges of radii as in Figure \ref{fig:per_dist}.  
Here we consider planets with $P < 50$ days and 
expand beyond the solar subset to \teff\ = 3600--7100 using the same cuts in \logg\ (4.0--4.9) and $\kp$ ($< 15$) 
to select bright main sequence stars.
We include only target stars for which photometric noise permits the detection of 2 \rearth planets, 
correcting for reduced detectability of small planets around the larger, hotter stars.
The occurrence of small planets (\rp\ = 2--4 \rearthe, orange histogram) rises substantially with decreasing \teff.  
The best-fit linear occurrence model for these small planets is shown as a red line.
The number of stars in each temperature bin is listed at the top of the figure.  
MK spectral types \citep{Allen2000} for main sequence stars are shown for reference.}
\label{fig:teff_dist}
\end{figure}

Figure \ref{fig:teff_dist} shows these occurrence measurements as a function of \teff.
Most strikingly, occurrence is inversely correlated with \teff\ for small planets with \rp\ = 2--4 \rearthe.  
Fitting the occurrence of these small planets in the \teff\ bins shown in Figure \ref{fig:teff_dist}, 
we find that a model linear in \teff, 
\begin{equation}
f(T_{\rm eff}) = f_0 + k_T \left(\frac{T_{\rm eff} - 5100\ \mathrm{K}}{1000\ \mathrm{K}}\right),
\label{eq:teff}
\end{equation}
fits the data well.  
Using linear least-squares, the best-fit coefficients are $f_0$ = \TeffEtaNaught\ and $k_T$ = \TeffGammaT\   
and the relation is valid over \teff\ = 3600--7100 K.  
We adopted a linear model because it is simple and provides a satisfactory fit with a reduced $\chi^2$ of \TeffChiSqReduced.
However, we caution that the occurrence measurements in the three coolest bins have relatively large errors and 
are consistent with a flat occurrence rate, independent of \teff.  
 
The occurrence of planets with radii larger than 4~\rearth does not appear to correlate with \teff\ (Figure \ref{fig:teff_dist}), 
although detecting such a dependence would be challenging given the lower occurrence of  
these planets and the associated small number statistics in our restricted sample.

\subsection{Sources of Error and Bias}

The correlation between the occurrence of 2--4 \rearth planets and \teff\ is striking.  
In this subsection we consider three possible sources of error and/or bias that could have spuriously produced this result.  
First, we rule out random errors in the occurrence measurements or in the stellar parameters in the KIC.  
Next, we consider a systematic bias in \rstar, but conclude that any such bias will be too small to cause the correlation.  
Finally, we consider a systematic metallicity bias as a function of \teff.
While we consider this unlikely, we cannot rule it out as the cause of the observed correlation.


\subsubsection{Random Errors}

One might worry that the fit to equation (\ref{eq:teff}) is driven by fluctuations due to small number statistics 
in the coolest temperature bins.  
The monotonic trend of rising planet occurrence from 7100 K to 4600 K is less clear for the two coolest bins with 
\teff\ = 3600--4600 K.
The coolest \teff\ bin, 3600--4100 K, contains only six detected planets and carries the largest uncertainty of any bin.  
The 4100--4600 K bin contains 13 detected planets.
As a test we excluded the hottest and coolest \teff\ bins and fit equation (\ref{eq:teff}) to the 
remaining occurrence measurements (4100--6600 K).  
The best-fit parameters were unchanged to within 1-$\sigma$ errors.

Next, we checked to see if random or systematic errors in stellar parameters 
could cause the correlation of 2--4 \rearth planet occurrence with \teff. 
The key stellar parameters from the KIC are \teff\ and \logg\ which have RMS errors of 135 K and 0.25 dex, respectively.  
Stellar radii carry fractional errors of 35\% RMS stemming from the \logg\ uncertainties.  

Using a Monte Carlo simulation, we assessed the impact of these random errors in the KIC parameters 
on the noted correlation. 
In 100 numerical realizations, we added gaussian random deviates to the measured \teff\ and \logg\ values for every star in the KIC.  
These random deviates, $\Delta$\logg\ and $\Delta$\teff,  
had RMS values equal to the RMS errors of their associated variables (135 K and 0.25 dex).  
Using the new \logg\ values we updated $\rstar$ for every star using 
$R_{\star\mathrm{,new}} = R_{\star\mathrm{,old}} 10^{\Delta\logg/2}$.
Planet radii, \rp, were updated in proportion to the change in \rstar\ for their host stars.
With each simulated KIC we performed the entire analysis of this section: 
we selected KIC stars that meet the \teff, \logg, and \kp\ criteria, divided those stars into 500 K subgroups, 
computed the occurrence of \rp\ = 2--4 \rearth planets in each \teff\ bin using the perturbed \rp\ values, 
and fit a linear function to the occurrence measurements in each \teff\ bin yielding $f_0$ and $k_T$.  
The standard deviations of the distributions of $f_0$ and $k_T$ from the Monte Carlo runs are 0.011 and 0.009, respectively.  
These uncertainties are nearly equal to the statistical uncertainties of $f_0$ and $k_T$ quoted above 
that are derived from the binomial uncertainty of the number of detected planets within each \teff\ bin.  
Thus, our quoted errors on $f_0$ and $k_T$ above probably underestimate the true errors by $\sim\sqrt{2}$.  
We conclude that the correlation between \teff\ and the occurrence of 2--4 \rearth planets 
is not an artifact of random errors in KIC parameters.  

\subsubsection{Systematic \rstar\ Bias?}

Potential systematic errors in the KIC parameters present a greater challenge than random errors.  
We assessed the impact of systematic errors by considering the null hypothesis---that the occurrence of 2--4 \rearth planets 
is actually independent of \teff---and determined how large the systematic error in \rstar(\teff) would have to be 
to produce the observed correlation of occurrence with \teff\ (equation \ref{eq:teff}).  
That is, systematic errors have to account for the factor of 7 increase in the occurrence of 2--4 \rearth 
planets between the \teff\ =  6600--7100 K and 3100--3600 K bins.
In this imagined scenario, the photometric determination of \logg\ in the KIC has a systematic error that is a function of \teff.  
This systematic error causes corresponding errors in \rstar\ and ultimately \rp\ that depend on \teff.  
We assumed that the power law radius distribution measured in Section \ref{sec:occ_rp} is independent of \teff\ and that 
it remains valid for $\rp < 2$ \rearthe.  
Then the systematic error in \rp\ would shift the bounds of planet radius in each \teff\ bin.  
That is, in the lowest \teff\ bin (3100--3600 K), while we intended to measure occurrence for planets with radii 2--4 \rearthe, 
we actually measured occurrence over a range of smaller radii, (2--4 \rearthe)/$S$, where the occurrence rate is intrinsically higher.  
Here, $S$ is a dimensionless scaling factor that describes the size of the systematic \rp\ error in the \teff\ = 3100--3600 K bin.
Similarly, for the \teff\ = 6600--7100 K bin we intend to measure the occurrence of 2--4 \rearth planets, 
but instead we measure the occurrence of planets with \rp\ = $S\cdot$(2--4 \rearthe) 
because of systematic errors in \rp(\teff) $\propto$ \rstar(\teff).  
Using the power law dependence for occurrence with \rp\ (equation \ref{eq:occ_rp}), we find that $S = (1/7)^{\alpha/2} = 6.2$ 
for the systematic error in \rp(\teff) to cause a factor of 7 occurrence error between the coolest and hottest \teff\ bins.  
A factor of 6.2 error in \rstar\ corresponds to a \logg\ error of 1.6 dex and is akin to mistaking a subgiant for a dwarf.  
Surely systematic errors in \rstar\ and \logg\ from the KIC are smaller than this.  
The KIC was constructed almost entirely for the purpose of selecting targets for the planet search by excluding evolved stars.  
\citet{Brown2011_kic} compared the \logg\ values from the KIC and LTE spectral synthesis of Keck-HIRES spectra 
and found that only one star out of 34 tested had a \logg\ discrepancy of greater than 0.3 dex (see their Figure 8).
We reject the null hypothesis and conclude that the strong correlation between the occurrence of 2--4 \rearth planets and \teff\ is real.  

\subsubsection{Systematic Metallicity Bias?}
\label{sec:occ_feh}

Another potential bias stems from the metallicity gradient as a function of height above the galactic plane \citep{Bensby2007,Neves2009}.
The \Kepler field sits just above the galactic plane, with a galactic latitude range $b = 6$--20 degrees.  
The most luminous and hottest stars observed by the magnitude-limited \Kepler survey are on average the most distant.  
Because of the slant observing geometry, these stars also have the greatest height above the galactic plane.
Likewise, the least luminous and coolest stars observed by Kepler are closer to Earth and only a small distance above the plane.  
Given that the average metallicity declines with distance from the galactic plane, 
one might expect that the hottest stars have lower metallicity, on average, than the coolest stars observed by \Keplere.

This hypothesis suggests a key test: does the occurrence of 2--4 \rearth planets depend on \feh?
Unfortunately we are not able to perform this test using stellar parameters from the KIC.  
While \teff\ values are accurate to 135 K (rms), \feh\ values are of poor quality.  
\citet{Brown2011_kic} found \feh\ errors of 0.2 dex (rms), and possibly higher due to systematic effects.  
Thus, the \feh\ values from the KIC are not helpful in testing the hypothesis that the occurrence of 2--4 \rearth planets 
depends on metallicity.

To get a sense of the size of the metallicity gradient as a function of \teff, 
we simulated our magnitude-limited observations of the \Kepler field 
using the Besancon model of the galaxy \citep{Robin2003}.  
This simulation produced a synthetic set of stars (with individual values of \teff, \logg, \feh, \mstar, etc.)\ based on the 
coordinates of the \Kepler field.  
We computed the median \feh\ for the seven \teff\ bins in Figure \ref{fig:teff_dist} and found, from coolest to hottest, 
\feh\ (median) = $-$0.02, $-$0.03, $-$0.03, $-$0.06, $-$0.07, $+$0.01, $+0.04$.   
The somewhat surprising upturn in metallicity in the two hottest \teff\ bins appears to be due to an age dependence with \teff; 
younger stars are more metal rich.  
The two hottest bins have a median age of 2 Gyr, while the five cooler \teff\ bins have median ages of 4--5 Gyr.  
We conclude based on this synthetic galactic model that \feh\ varies by perhaps $\sim$0.1 dex over our \teff\ range and that the 
dependence need not be monotonic due to age effects.  

It is also worth considering how large of an \feh\ gradient is needed to increase \textit{giant} planet occurrence 
by a factor of seven.  
Clearly, occurrence trends for jovian planets and 2--4 \rearth planets need not be similar, 
but these larger planets offer a sense of scale than may be relevant for smaller planets.  
For giant planets, \citet{Fischer2005} found that occurrence scales as $\propto 10^{2.0\mathrm{[Fe/H]}}$, 
while \citet{Johnson2010} found $\propto 10^{1.2\mathrm{[Fe/H]}}$, after accounting for the occurrence dependence on \mstar.  
These scaling relations suggest that \feh\ gradients of 0.4--0.7 dex are needed to affect a factor of seven change in occurrence.  
A metallicity change of only $\sim$0.1 dex among 2--4 \rearth planet hosts seems unlikely to change planet occurrence 
by the amount we observed.  Further, if the occurrence of such planets depends so sensitively on \feh, 
it seems likely that Doppler surveys of them would have detected this trend among the $\sim$30 RV-detected planets 
with \msini\ $<$ $M_\mathrm{Neptune}$.

The possibility that increased metallicity correlates with increased 2--4 \rearth planet occurrence 
contradicts tentative trends of low-mass planets observed by Doppler surveys.
\citet{Valenti10} noted that among the host stars of Doppler-detected planets, those stars with \textit{only} 
planets less massive than Neptune are metal poor relative to the Sun.   
This tentative threshold is intriguing, but it only shows that the distribution of \textit{detected} planets has an apparent \feh\ threshold, 
not that the occurrence of these planets depends systematically on \feh.
To interpret the threshold physically, one needs to check for metallicity bias in the population of Doppler target stars.  


\section{Planet Density}  
\label{sec:density}

It is tempting to extract constraints on the densities of small
planets by comparing the distribution of radii measured by \Kepler to
the distribution of minimum masses (\msini) measured by
Doppler-detected planets from surveys of the Solar neighborhood
\citep{Cumming08, Howard2010b}.  
This effort may be partially compromised by the
different populations of targets stars, despite our efforts to select stars with similar 
\logg\ and \teff\ distributions.  
The \Kepler target stars are typically $\sim$50--200 pc above the Galactic plane while Doppler
target stars reside typically within 50 pc of the Sun near the plane.
Indeed in Section \ref{sec:occ_per} we saw that the hot Jupiter occurrence was 2.5 
times lower in the \Kepler survey than in the Doppler surveys,
suggesting a difference in stellar populations, possibly related to
the decline in metallicity with Galactic latitude and/or differing \teff\ distributions.  
Nonetheless, one should not ignore the opportunity to search for information from
combining the \Kepler and Doppler planet occurrences, with caveats
prominently in mind.

We first consider known individual planets that have measured masses, radii, and implied bulk densities.  
Placing these well-measured planets on theoretical mass-radius
relationships \citep[e.g.,][]{Valencia2006,Seager2007,Sotin2007,Baraffe2008,Grasset2009} 
provides insight into the range of compositions encompassed by the detected planets.  Our goal is to 
complement these few well-studied cases with statistical constraints on the planet density distribution.

\subsection{Known Planets}

\begin{figure}
\includegraphics[width=0.45\textwidth]{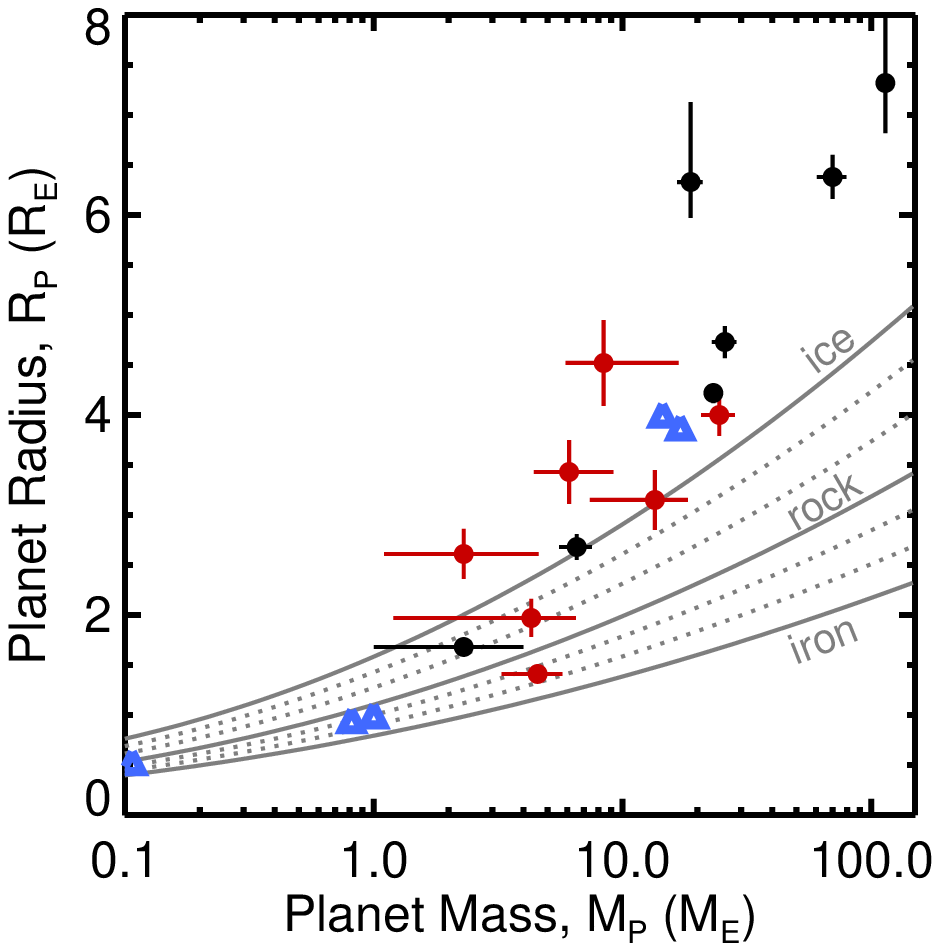}
\includegraphics[width=0.45\textwidth]{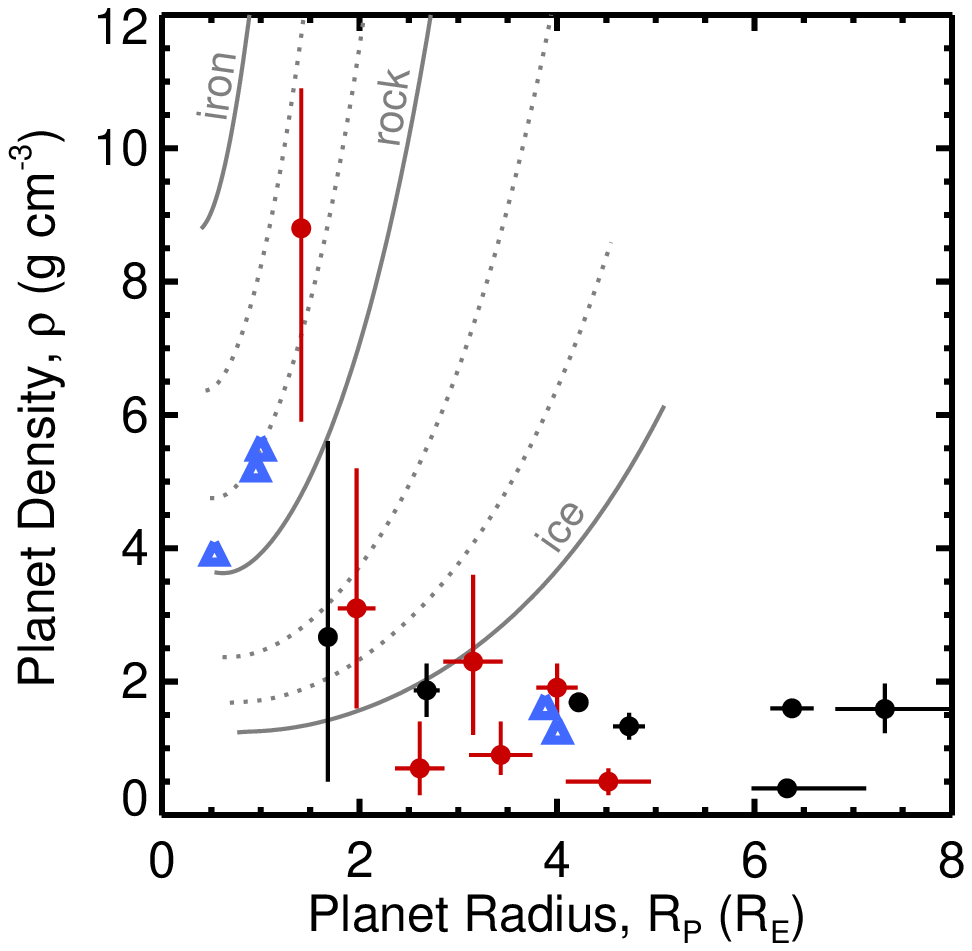}
\caption{Planet masses, radii, and densities for $\rp < 8$ \rearth and $M_{\rm p} > 0.1$ \mearthe.
The upper panel shows radius as a function of mass and the lower panel shows density as a function of radius.
Solar system planets (Mars, Venus, Earth, Uranus, and Neptune) are depicted as blue triangles.
Extrasolar planets (filled circles) are colored red for \Kepler discoveries and black for discoveries by other programs.  
Solid gray lines indicate the densities of solid planets composed of pure ice, pure rock, and pure iron using the 
\citet{Fortney2007_errat,Fortney2007} models.  
Dotted gray lines show the densities of admixture compositions (from bottom to top in lower panel):
67/33\% ice/rock, 33/67\% ice/rock, 67/33\% rock/iron, 33/67\% rock/iron.  
}
\label{fig:density_dist_known}
\end{figure}

We begin by considering the known planets with $\rp < 8$ \rearth and 
$M_{\rm p} > 0.1$ \mearthe.  
This range of parameters selects planets smaller than Saturn and as large or larger than 
Mars.  Figure \ref{fig:density_dist_known} shows all 
such planets with good mass and radius measurements from our solar 
system and other systems.  Theoretical calculations of Kepler-10b
\citep{Batalha2011} based on its mass and radius (4.5 \mearth and 1.4
\rearthe) suggest a rock/iron composition with little or no water.
Corot-7b has a radius of 1.7 \rearth \citep{Leger09}.  
\citet{Queloz2009} measured a mass of 4.8 \mearth for this planet, 
implying a density of 5.6 g\,cm$^{-3}$ and a rocky composition.  
However, the mass and density have remained controversial.  
Independent mass determinations based on the same spot-contaminated Doppler data
yield masses that vary by a factor of 2--3 \citep{Pont2011, Hatzes2010, Ferraz-Mello2010}.  
We adopt the mass estimate of 1--4 \mearth from \citet{Pont2011}, which implies
a wide range of possible compositions and also marginally favors a
water/ice-dominated planet.  GJ~1214b is a less dense super-Earth
orbiting an M dwarf.  The planet has been modeled as a solid core
surrounded by H/He/H$_2$O and may be intermediate in composition between
ice giants like Uranus and Neptune and a 50\% water planet
\citep{Nettelmann2010}.  
The discovery of the six co-planar planets
orbiting Kepler-11 added five planets with measured masses (from
transit-timing variations) to Figure
\ref{fig:density_dist_known} \citep{Lissauer2011}.  The remaining
exoplanets in Figure \ref{fig:density_dist_known} all have masses 
greater than Neptune's (17 \mearthe) and densities less than 2 
g\,cm$^{-3}$: Kepler-4b \citep{Borucki2010_Kepler-4}, Gl~436b
\citep{Maness2007,Gillon2007, Torres2008}, HAT-P-11b \citep{Bakos09},
HAT-P-26b \citep{Hartman2011}, Corot-8b \citep{Borde2010}, HD~149026b
\citep{Sato2005,Torres2008}.

Figure \ref{fig:density_dist_known} shows that among known planets
their radii increase with planet mass faster than do any of the
theoretical curves representing solid compositions of iron, rock, or
ice.  This rapid increase in radius with mass suggests that planets of
higher mass contain larger fractional amounts of H/He gas.  The slope
increases markedly for masses above 4.5 \mearthe, indicating that above
that planet mass the contribution of gas is common, even for these
close-in planets.  Apparently planets above 4.5 \mearth are rarely
solid.  We suspect that for planets orbiting beyond 0.1 AU where
collisional stripping of the outer envelope is less energetic and
common, the occurrence of gaseous components will be greater.

\citet{Fortney2007} modeled solid exoplanets composed of pure water 
(``ice''), rock (Mg$_2$SiO$_4$), iron, and binary admixtures.  Their 
models include no gas component and are shown as gray lines in Figure 
\ref{fig:density_dist_known}.  Adding gas to any of the models 
increases \rp\ and decreases $\rho$ \citep{Adams2008}.  Thus planets below and to the 
right of the ice contour (Figure \ref{fig:density_dist_known}, lower 
panel) have low densities due to a gas component.  Planets above the 
ice contour contain increasing fractions of rock and iron, depending 
on the specific system.  Compositional details matter greatly for 
specific systems, but for our simple purpose we make the crude 
approximation that planets with $\rp \lesssim 3$ \rearth that have 
$\rho \gtrsim 4$ g\,cm$^{-3}$ are composed substantially of refractory materials 
(usually rock in the form of silicates and iron/nickel).  
These planets may have some water and gas, but those components do not 
dominate the planet's composition as they do for Uranus, Neptune, and 
larger planets.

\subsection{Mapping Kepler Radii to Masses}

The Eta-Earth Survey measured planet occurrence as a function of \msini\ in a volume-limited
sample of 166 G and K dwarfs using Doppler measurements from Keck-HIRES.  
The stars have a nearly unbiased metallicity distribution and are
chromospherically quiet to enable high Doppler precision.
In all, 35 planets were detected around 24 of the 166 stars, 
including super-Earths and Neptune-mass planets \citep{Howard09a,Howard2010c,Howard2011a}.
Correcting for inhomogeneous  sensitivity at the lowest planet masses, 
\citet{Howard2010b} measured increasing planet occurrence with decreasing mass over 
five planet mass domains, \msini\ = 3--10, 10--30, 30--100, 100--300, 300--1000 \mearthe, 
spanning super-Earths to Jupiter-mass planets.  
This study was restricted to planets with $P < 50$ days.

We mapped the planet radius distribution from \kepler (Figure \ref{fig:rp_per}, including planets down to 1 \rearthe) 
onto mass (\msini) using toy density functions, $\rho(\rp)$.  
These single-valued functions map all planets of a particular radius, \rp, onto 
a planet mass $M_{\rm p} = 4\pi \rho(\rp)\rp^3/3$.  
Of course, real planets exhibit far more diversity in radii for a given mass 
owing to different admixtures of primarily iron/nickel, rock, water, and gas.  
Nevertheless, the models allow us to check if average masses associated with \Kepler 
radii are consistent with Doppler measurements.  

As part of this numerical experiment we converted $M_{\rm p}$ to \msini\ for each simulated planet using random orbital orientations 
(inclinations $i$ drawn randomly from a probability distribution function proportional to \sini.)
Our simulated \msini\ distributions account for the transit probabilities of planets detected by \Kepler 
and the detection incompleteness for planets with small radii.  
That is, the simulated \msini\ distributions reflect the true distribution of planet radii (Section \ref{sec:occ_rp}).

\begin{figure*}
\includegraphics[width=1.0\textwidth]{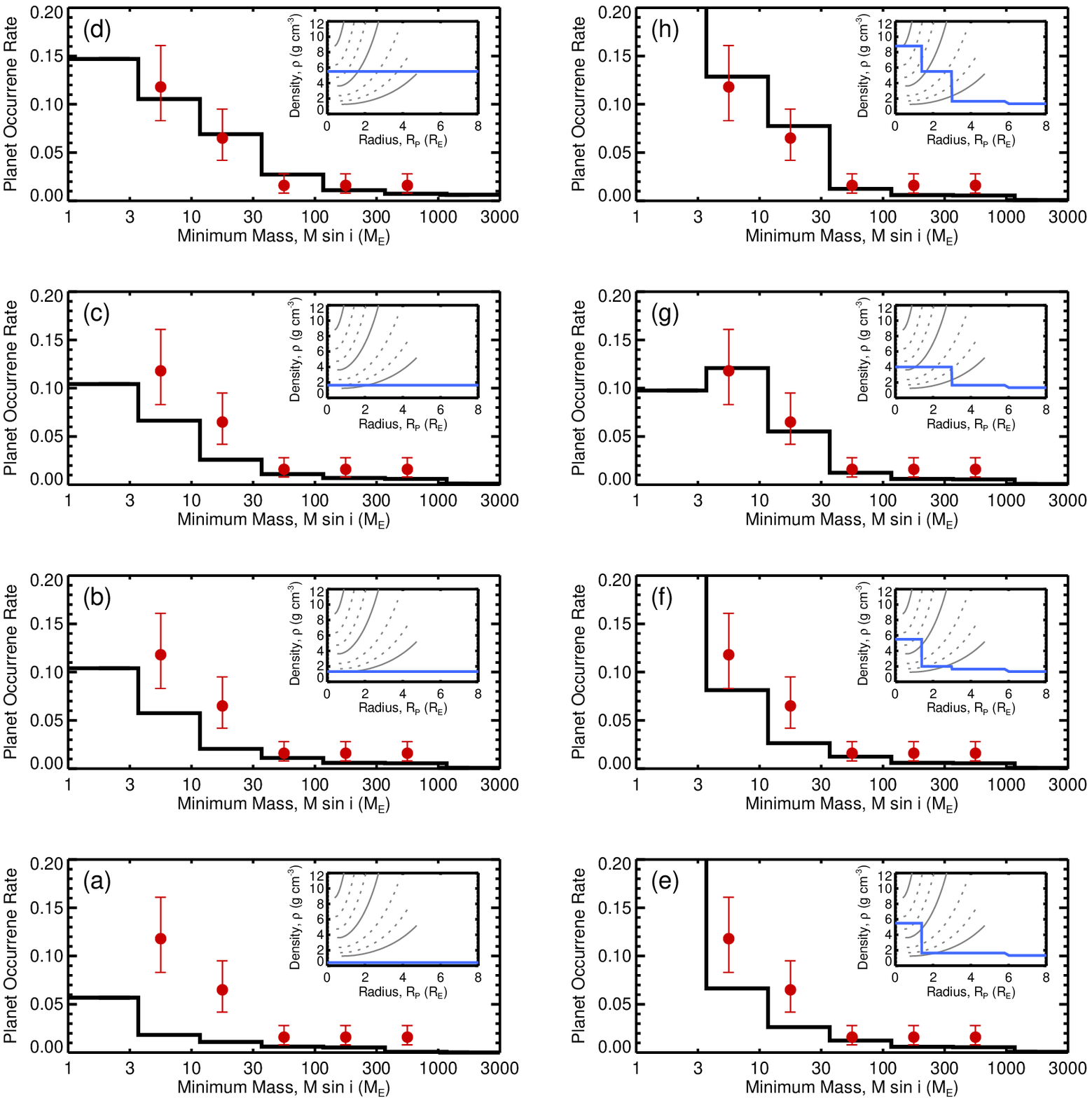}
\caption{Predicted mass distributions (\msini, black histograms) based on planet radii measured by 
\Kepler and hypothetical planet density functions (blue lines in inset panels).
The \citet{Fortney2007_errat,Fortney2007} theoretical density curves for solid planets 
from Figure \ref{fig:density_dist_known} are plotted as solid and dotted gray lines in each inset panel.
Planet occurrence measurements as a function of \msini\ from the 
Eta-Earth Survey \citep{Howard2010b} are shown as red filled circles.  
Panels in the left column show the mass distributions resulting from toy constant-density models.  
From bottom to top (panels a--d), all planets have densities of 0.4, 1.35, 1.63, and 5.5 g\,cm$^{-3}$, independent of radius, 
in analogy with the densities of Earth, Neptune, Jupiter, and HAT-P-26b.
In the right column (panels e--h) density increases with decreasing planet radius, as depicted by the inset density functions.   
Density functions that increase above $\sim$4.0 g\,cm$^{-3}$ for planets with $\rp \lesssim 3$ \rearth 
yield greater consistency between the Eta-Earth Survey and \Keplere.
}
\label{fig:density_dist}
\end{figure*}

Figure \ref{fig:density_dist} shows simulated \msini\ distributions assuming several toy density functions.  
These distributions are binned in the same \msini\ intervals as in the \citet{Howard2010b} study.
In the left column $\rho(\rp) = \rho_0$, where $\rho_0$ is a constant.  
From bottom to top, we considered four densities, $\rho_0$ = 0.4, 1.35, 1.63, and 5.5 g\,cm$^{-3}$ 
(the bulk densities of HAT-P-26b, Jupiter, Neptune, and Earth).
We are most interested in the densities of small planets so we make comparisons in the two lowest mass 
bins for which Eta-Earth Survey measurements are available, \msini\ = 3--10 and 10--30 \mearthe.
In these bins, the predicted occurrence from \Kepler is too small by 1.5--2$\sigma$ compared with 
the Eta-Earth Survey measurements for the three lowest constant density models, $\rho_0$ = 0.4, 1.35, and 1.63 g\,cm$^{-3}$.  
\Kepler predicts fewer small planets than the Eta-Earth Survey measured.  
The simulated \msini\ distribution matches the observed
\msini\ distribution well for an assumed density, $\rho$ = 5.5 g\,cm$^{-3}$.
While this model is clearly unphysical when extended over the entire radius range, 
consistency in the two low-mass bins suggests that the small planets have higher densities.  

We explored slightly more complicated density functions in the right column of Figure \ref{fig:density_dist}.  
These functions are piece-wise constant density models, 
with density rising to 4.0, 5.5, and 8.8 g\,cm$^{-3}$ 
for small radii, as depicted in the sub-panels of Figure \ref{fig:density_dist}.  
(Kepler-10b has a density of 8.8 g\,cm$^{-3}$; \citealt{Batalha2011}.) 
We find the greatest consistency between the synthetic and measured 
mass distributions for two density models. 
One (model h) is shown in the upper right panel of  Figure \ref{fig:density_dist} which has 
$\rho$ = 8.8, 5.5, 1.64, 1.33 g\,cm$^{-3}$ for \rp\ = 1--1.4,
1.4--3.0, 3.0--6.0, and $>$6.0 \rearthe, respectively.   This model
has a high density (8.8 g\,cm$^{-3}$) for the smallest planets but
successively smaller densities for larger  planets, approximately consistent with
the densities of known planets in Figure \ref{fig:density_dist_known}.  The
other successful model (g) has a density of 4 g\,cm$^{-3}$ for the
smallest planets, with declining densities for larger planets,
qualitatively similar to the previous model (h).  This model (g) also
yields a predicted distribution of \msini\ that agrees well with the
observed distribution of \msini.  Thus it too is viable.  Both
successful models, g and h, are characterized by a high density for
the smallest planets of 1--3 \rearthe.  We tried a variety of piecewise constant density functions 
and found that all models that achieved consistency 
($<1\sigma$ difference in the 3--10 and 10--30 \mearthe bins) have 
$\rho \gtrsim 4$ g\,cm$^{-3}$ for $\rp \lesssim 3$ \rearthe.  

\subsection{Conclusions}

The mapping of radius to mass offers circumstantial evidence that a 
substantial population of small planets detected by \Kepler have high densities.  
Rocky composition for the smallest planets supports the core-accretion model of planet formation 
\citep{Pollack1996, Lissauer2009, Movshovitz2010}.  
But we caution again that the stellar populations of the \Kepler and Doppler surveys may be quite different.
Planet multiplicity also makes this an especially challenging comparison.  
We computed the simulated \Kepler mass distributions (black histograms in Figure \ref{fig:density_dist}) 
based on occurrence measured as the average number of planets per star while the 
Doppler results from the Eta-Earth Survey (red points in Figure \ref{fig:density_dist}) 
computed occurrence as the fraction of stars hosting at least one planet in the specified \msini\ interval.  
This difference is based on intrinsic limitations of each approach.  
To infer the fraction stars with at least one planet from a transit survey requires an assumption 
about the mutual inclinations \citep{Lissauer2011b}.  
For Doppler surveys, it is significantly easier to determine if a particular star has 
at least one planet down to some specified mass limit, but it is much more 
difficult to be sure that all planets orbiting a star have been detected down to that same mass limit \citep{Howard2010b}.  
Finally, we note that no planets at the extreme of our proposed high density regime 
(\rp\ $\sim$ 3 \rearth and $\rho \sim 4$ g\,cm$^{-3}$) have been detected (Figure \ref{fig:density_dist}).  
To date all detected planets with \rp\ $>$ 2 \rearth have $\rho < 2$ g\,cm$^{-3}$.
We conclude that while this technique offers qualitative support for rising density with decreasing planet size, 
in practice extracting firm quantitative conclusions is difficult because of the 
intrinsic differences between Doppler and transit searches.


\section{Discussion}
\label{sec:discussion}

\subsection{Methods}

We have attempted to measure pristine properties of planets that can
be compared with, and can inform, theories of the formation, dynamical
evolution, and interior structures of planets.  We have built upon the
unprecedented compendium of over 1200 planet candidates found by the
historic \Kepler mission \citep{Borucki2011}.  One goal here was to
measure planet occurrence---the number of planets per star having
particular orbital periods and planet radii---by minimizing the
deleterious effects of detection efficiencies that are a function of
planet properties, notably radius and orbital period.  

Our treatment of the vast numbers of target stars and transiting
planet candidates involved careful accounting of two important
effects.  First, only planets whose orbital planes are nearly aligned
to \Keplere's line of sight will transit their host star, leaving many
planets undetected.  We applied the standard geometrical correction
for the small probability, $R_{\star}/a$ in equation (\ref{eq:eta}),
that the orbital plane is sufficiently aligned to cause a transit.  In
counting planets, we assumed that for each detected planet candidate
there are actually $a/R_{\star}$ planets, on average, at all
inclinations.  Second, only planets whose transits produce photometric
signals exceeding some signal-to-noise threshold will be reliably
detected.  For each possible planet radius and orbital period, we
carefully identified the subset of the \Kepler target stars 
{\it a  priori} around which such planets could be detected with high
probability.  We adopted a threshold SNR of 10 for the transit signal
in a single 90~day quarter of data, thereby limiting both the target
stars and the planet detections with this SNR threshold.  To be
included, a target star must have a radius and photometric noise that
allowed a planet detection with SNR $>$ 10, i.e. a transit depth 10
times greater than the uncertainty in the mean depth from noise.  Such
restricted target stars offer a high probability that planets will be
detected. 

We further selected \Kepler target stars having a specific range of
\teff, \logg, and brightness to ensure a well defined sample of stars.
We consider only bright target stars ($\kp < 15$).  We ignore all
other Kepler target stars and their associated planets.  Remarkably,
this {\it a priori} selection of \Kepler target stars immediately
yields a sample of only $\sim$58,000 stars (and fewer when accounting
for requisite photometric noise), not the full 156,000 stars.  For
most of the paper, we restricted the sample to main sequence G and K
stars (\logg\ = 4.0--4.9, \teff\ = 4100--6100 K) to permit comparison
with similar Sun-like stars in the Eta-Earth Survey.  This selection
of \kepler target stars for a given planet radius and orbital period
crucially leaves only a subset of stars in the ``sub-survey'' for
those planet properties.  Importantly, for planets with small radii (near
2 \rearthe) and long periods (near 50 days), only some 36,000--49,000
stars are amenable to detection of such difficult-to-detect planets,
as shown in the the annotations in the lower left corners of the cells
in Figure \ref{fig:rp_per}.  By counting planets and dividing by the
number of appropriate stars that {\it could} have permitted their
secure detection we computed the planets per star for a specific
planet radius and orbital period (within a specified delta in each
quantity).

\subsection{Comparison with \citet{Borucki2011}}

It is worth describing the differences between this paper and \citet{Borucki2011}
resulting from differing goals and methods.  
The primary propose of \citet{Borucki2011} was to summarize the
results of the \Kepler observations and to act as a guide to the
tables of data.  In doing so they tripled the number of known planets
\citep[even when allowing for a false positive rate of $\sim$5\%;][]{Morton2011}.
\citet{Borucki2011} considered the \textit{number distributions}
of all planets detected by \Keplere, independent of the properties of
their host stars (\teff, \logg, \kp, $\sigma_{\rm CDPP}$).  
They also computed the ``intrinsic frequencies'' of planetary candidates, a
close cousin of our planet occurrence measurements, and plotted these
frequencies as a function of \teff.  

The results in this paper are derived directly from the planets announced in \citet{Borucki2011} 
and from stellar parameters in the KIC \citep{Brown2011_kic}.  
We measure the occurrence distributions of 
planets orbiting bright, main sequence G and K stars, which represent 
only a third of the stars observed by \Kepler and considered in \citet{Borucki2011}.
Our desire for high detection completeness compelled us to consider only robustly 
detected planets satisfying $\rp > 2$ \rearthe, $P <  50$ 
days, SNR $>$ 10 in 90 days of photometry, and stars with $\kp < 15$.  
This selection of stars and planets facilitated comparison with the Eta-Earth Survey \citep{Howard2010b}, 
which focused on the Doppler detection of planets orbiting G and K dwarfs with $P < 50$ days.  
In this paper we measured the detailed patterns of planet occurrence as a function of 
\rp, $P$, and \teff, only for that subset of stars and interpreted these distributions in the context 
of planet formation, evolution, and composition.

\citet{Borucki2011} chose to compute intrinsic frequencies in small domains 
of semi-major axis and planet radius while we work in a space of orbital periods and planet radii.  
There are trade-offs with these choices.  
We chose to work in period space because \Kepler directly measures orbital periods  
and translating to semi-major axes requires either assumed stellar masses or radii.
On the other hand, by working in small domains of semi-major axis, \citet{Borucki2011} 
compensate for this by considering the range of orbital periods and transit durations 
that contribute to each domain for the range of masses and radii among the target stars.  
In this paper we applied a binary detection criterion of SNR $>$ 10 for 90 days of photometry 
(approximately one quarter).  
\citet{Borucki2011} adopted a detection criterion of SNR $>$ 7 for the 136 days of Q0--Q2, 
with corrections for the probability of low SNR detections (e.g.\ 7-$\sigma$ detections are 
only recognized 50\% of the time).

\subsection{Patterns of Planet Occurrence}

Figure \ref{fig:rp_per} shows graphically some of the key features of
close-in planet occurrence.  The number of planets per star varies by
three orders of magnitude in the radius-period plane (Figure
\ref{fig:rp_per}) that spans periods less than 50 days and planet
radii less than 32 \rearthe.  Planet occurrence increases toward smaller
radii (see Figure \ref{fig:rp_dist}) down to our completeness limit of 2
\rearthe, with a power law dependence given by 
$\mathrm{d}f(R)/\mathrm{d}\log_{10} R = k_R R^{\alpha}$
where $\mathrm{d}f(R)/\mathrm{d}\log_{10} R$ is the number of planets per star, for planets with 
$P < 50$ days in a $\log_{10}$ radius interval centered on $R$ (in
\rearthe), $k_R$ = \kRPL\ , and $\alpha$ = \alphaRPL.  This is a
remarkable result, showing that from planets larger than Jupiter to
those only twice the radius of Earth planet occurrence rises rapidly
by nearly two orders of magnitude.  This rise with smaller size is consistent
with, and supports the measured rise of, the planet occurrence with
decreasing planet {\it mass} found by \cite{Howard2010b}.  
The increased occurrence of small planets seen in both studies 
supports the core-accretion theory for planet formation \citep{Pollack1996}.

Planet occurrence also increases with orbital period
(Figure \ref{fig:per_dist}) in equal intervals of $\log P$ as $f(P) =
k_P P^{\beta} \left(1-e^{-(P/P_0)^\gamma}\right)$ with coefficients
that all depend on planet radius, and both $\beta$ and $\gamma$ being
positive.  This functional form traces the steep rise in planet
occurrence near a cut-off period, $P_0$.  Below $P_0$ planets are
rare, but for longer periods the planet occurrence distribution rises
modestly with a power law dependence.  We find that $P_0$ and $\gamma$
(which governs how steep the occurrence fall-off is below $P_0$)
depend on planet radius.  The smaller planets, \rp\ = 2--4 \rearth,
have $P_0 \sim 7$ days, while larger planets have a $P_0 \sim 2$ days.
Further, $\gamma$ is larger for planets with $\rp > 4$ \rearth making
the fall-off in planet occurrence more abrupt below $P_0$.  The trends
suggest that the mechanisms that caused the planets to migrate and
stop at close orbital distances depend on planet size.  
Alternatively, if a substantial number of small close-in planets formed by \textit{in situ} accretion, 
then our measurements trace the contours of this process \citep{Raymond2008}.

This period dependence of planet occurrence seems to contradict the
results from Doppler surveys of exoplanets for which we find a pile-up
of planets at periods of 3 days and a nearly flat distribution of
planets for longer periods, out to periods of 1 yr \citep{Wright2009}.
The key difference is that \Kepler is sensitive to much smaller
planets (in radius and mass) than were Doppler surveys, especially
beyond 0.1 AU.  To be sure, \Kepler suffers a geometrical decline in
detectability as $R_{\star}/a$ but we have corrected for this trivially.  
Such a correction is more difficult for Doppler surveys that have less 
uniform detectability from star to star.

Another difference in the period distributions between \Kepler and
Doppler surveys is in the pile-up of hot Jupiters at orbital periods
near 3 days (Figures~\ref{fig:rp_per} and \ref{fig:per_dist}).  The
\Kepler detected planets show a pile-up, but it is modest, almost not
significant, while for single planets in Doppler surveys the pile-up is
a factor of three above the background occurrence at other periods
\citep{Marcy08,Wright2009}.  This different planet occurrence for hot
Jupiters appears to be real, and may be due to fewer metal-rich stars
in the \Kepler sample that are located 50--200 pc above the Galactic
plane, or different stellar mass distributions in the magnitude-limited 
and volume-limited surveys.  
The \Kepler field has a greater admixture of thick disk stars
(that are metal poor with [Fe/H] $\approx$ $-$0.5) to thin disk stars
than do the Doppler target stars.  Other photometric surveys have
noted that hot Jupiter occurrence appears to vary with stellar
population.  \citet{Gilliland2000} found no planets in a
\textit{Hubble Space Telescope} survey of the globular cluster 47
Tucanae and estimated a hot Jupiter occurrence that is an order of
magnitude lower than in the solar neighborhood.  Similarly,
\citet{Weldrake2008} found no planets in the $\omega$~Centauri
globular cluster and found the occurrence of hot Jupiters ($P = 1$--5
days) to be less than 0.0017 planets per star.  \citet{Gould2006}
found an occurrence of $0.003^{+0.004}_{-0.002}$ hot Jupiters per star
for $P = 3$--5 days, based on the magnitude-limited OGLE-III survey in
the bulge of the Galaxy, and are compatible with our results from
\Keplere, \OccPtenReightthirtytwoMag\ planets per star for \rp = 8--32 \rearthe, 
$P < 50$ days, and $\kp < 16$.

We further find that planets larger than 16 \rearth (1.5
$R_{\mathrm{Jup}}$) are extremely rare.  Such inflated planets are
also rare among transiting planets detected from the ground (see,
e.g., the mass-radius diagram for gas giant planets in \citet{Bakos2010_HAT2023}).  
For several Gyr-old planets, theoretical
mass-radius curves show a maximum near $\rp \approx 13$ \rearth $\approx$ 1.2
$R_{\mathrm{Jup}}$ \citep{Fortney2007}.  
Larger planets are typically young or close-in and inflated by one of several 
proposed mechanisms \citep[e.g.,][]{Batygin2010, Laughlin2011,Burrows2007b}.

We also note some interesting morphology in the two-dimensional
occurrence domain of planet radius and orbital period (Figure \ref{fig:rp_per}).  
There is a ridge of higher planet occurrence for super-Earths and Neptunes,
similar to that identified in \cite{Howard2010b}.  The ridge appears
to be diagonal when plotting either $M_{\rm p}$ or \rp\ vs.\ $P$ extending from a period and
radius of 3 days and 2 \rearth (lower left) to a period and radius of
50 days and 4 \rearthe.  This ridge can be seen by direct inspection of
the Figure \ref{fig:rp_per}, both by the density of the dots and by
the colors.  The upper envelope of red boxes (indicating high planet
occurrence) extends along a diagonal from lower left to upper right.
This ridge conveys some key information about the formation and
perhaps dynamical evolution or migration of the 2--4 \rearth planets.

The paucity of close-in Neptune-mass planets (\msini\ = 10--100
\mearthe, $P < 20$ days) seen in \citet{Howard2010b} is not as clearly
visible in the Kepler data.  In particular, the ``top'' of this desert
(\msini\ = 100 \mearthe, or the radius equivalent) is not as clear.  A
further study of \Kepler stars to fainter magnitudes of \kp\ = 16 may
shed light on this desert.  The overall planet occurrence for GK stars
and periods less than 50 days, listed in Table 3, shows that planets
of 2--4 \rearth  is \OccPfiftyRtwofour\ planets per star.  This agrees well with the
planet occurrence of 3--30 \mearth planets found by \cite{Howard2010b}
of $15^{+5}_{-4}$\%.  The planet occurrence for all planet radii from 2--32 \rearth is
only 16.5\%, again in agreement with \cite{Howard2010b} and
\cite{Cumming08}.  We find little support for the suggestion of planet
occurrences of super-Earths and Neptunes (\msini\ = 3--30 \mearthe) of
30\% $\pm$ 10\% \citep{Mayor09a} for $P < 50$ days.
 
We also measured planet occurrence as a function of \teff\ of the host
star, a proxy for stellar mass.  For the smallest planets, 2--4
\rearthe, the results show a nearly linear rise in planet occurrence
with smaller stellar mass.  One may wonder if this rise might be
caused by some systematic error due to poor values of \teff\ or \rstar\ in the \Kepler Input
Catalog.  Such a systematic error seems nearly impossible, as the KIC
values of \teff\ are accurate to 135 K (RMS) and in any case the
\teff\ values certainly vary monotonically with the true value of
\teff\ even if one imagines some large systematic error in the KIC
values of \teff.  Thus the increase in planet occurrence with smaller
\teff\, and hence smaller stellar mass, appears to be real.  
Again, we emphasize that the SNR = 10 criterion for a \kepler target star to be
included in our survey implies that the detection efficiency is close to
unity for all stars, from 7100 K to 3600 K, for $R > 2$ \rearthe.  
Examination of Figure \ref{fig:teff_dist} shows that even if one ignores the coolest
and hottest stars, the increase of planet occurrence
persists robustly.  Thus it appears that the number of planets per
star increases by a factor of seven from stars of 1.5 \msun to stars
of 0.4 \msun (\teff\ = 7100--3600K), with all of that \teff\ dependence
coming from the smallest planets, 2-4 \rearthe.  This high occurrence of close-in small
planets around low mass stars represents significant information about
the formation mechanisms of planets of 2--4 \rearthe.  

We considered the possibility that this correlation is due to a systematic metallicity bias that depends on \teff.  
That is, cool stars are relatively nearby, close to the galactic plane, and have higher metallicities, 
while hot stars are on average more distant, at greater heights above the galactic plane, and 
have lower metallicities.  
In this scenario, low metallicity is the driving force behind lower planet occurrence at higher \teff.  
Using the Besancon galactic model, we estimate that metallicities may vary by $\sim$0.1 dex as a function of \teff, 
but the dependence need not be monotonic because of the median age varies with \teff.
It would be remarkable if such a modest difference in metallicity could cause a factor of seven difference 
in close-in planet occurrence.  
Unfortunately, due to the poor \feh\ measurements in the KIC, we are unable to measure the occurrence of 
planets as a function of \feh.
Note, however, that either result has profound implications for planet formation: 
the occurrence of 2--4 \rearth planets depends strongly on stellar properties, \teff\ or \feh.

Sub-Neptune-size and jovian planets appear to have opposite trends in occurrence as a function of \mstar.  
We showed that the occurrence of 2--4 \rearth planets 
decreases by a factor of seven with \mstar\ over $\sim$0.4--1.5 \msun (\teff\ = 3600--7100 K). 
\cite{Johnson2010} measured the occurrence of giant planets as a function of \mstar\ and [Fe/H] 
and found a positive correlation with both quantities.  
That is, the occurrence of giant planets \textit{increases} with increasing \mstar\ 
over the range $\sim$0.3--1.9 \msune.
Their study considered only giant planets that produce $K > 20$ \ms Doppler signals and orbit within 2.5 AU.  
Subgiants with \mstar = 1.4--1.9  \msun have the highest rate of giant planet occurrence in their study.  
However, most of these planets orbit at $\sim$1--2 AU, with almost no planets inside of $P$ = 50 day orbits \citep{Bowler2010}.  
Close-in planets of all sizes larger than 2 \rearth appear to be rare around the most 
massive stars accessible to transit and Doppler surveys.

\subsection{Planet Formation}
 
Population synthesis models of planet formation by core accretion 
simulate the growth and migration of planet embryos embedded in a
proto-planetary disk of gas and dust.  Among their key predictions is the 
distribution of planet mass or radius as a function of
orbital distance.  Early versions of these models \citep{Ida04a,
Alibert2005, Mordasini2009} were tuned to match the distribution of
giant planets detected by RV \citep{Cumming08,Udry2007} by decreasing
the rate of Type I migration compared to theoretical predictions
\citep{Ward1997,Tanaka2002}.  The simulations predicted that planet
occurrence rises with decreasing planet mass.  But most of the low-mass
planets resided in orbits near or beyond the ice line at $\sim$2--3 AU.
These models also robustly predicted a ``planet desert'', a region of
parameter space nearly devoid of planets.  Planets with \msini\ $\approx$
1--20 \mearth and $a \lesssim 1$ AU were predicted to be extremely
rare because producing such planets requires the gas disk to
dissipate while one of two faster processes were happening, Type II
migration or run-away gas accretion.  Meanwhile, the models predicted
that planets with masses above the
desert, $M > 20$\mearthe, but residing inside of $\sim$1 AU would
exhibit a nearly constant distribution with mass.

\citet{Howard2010b} demonstrated that the observed distribution of
close-in planets ($P < 50$ days) exhibited quite different properties
from those predicted by population synthesis.  The predicted planet desert is
actually populated by the highest planet occurrence of any region of
the mass-period parameter space yet probed (the ``ridge'' noted above).  The planet
mass function rises steeply with decreasing planet mass, in
contradiction to the expected nearly constant distribution with mass outside of the
desert.  From \Keplere, we also see many planets populating
the predicted desert (Figure \ref{fig:rp_per}) and a planet radius
distribution that rises steeply with decreasing planet size (tracking
the mass distribution).  The latest versions of the population
synthesis models \citep{Ida_Lin_2010, Alibert2011} offer improvements
including non-isothermal treatment of the disk
\citep{Paardekooper2010} and multiple, interacting planet embryos per
simulation.  But they still predict a planet desert (albeit partially
filled in).  The contours of planet occurrence in Figure
\ref{fig:rp_per} offer rich detail to which future refinements of
these models can be tuned.  Alternatively, the distribution of
observed planets may be strongly shaped by processes that take place
after the gas clears, namely planet-planet scattering
\citep[e.g.,][]{Ford2005, Ford2008, Chatterjee2008, Raymond2009},
secular and resonant migration \citep[e.g.,][]{Lithwick_Wu_2011,
Wu_Lithwick_2011}, and planetesimal migration and growth
\citep[e.g.,][]{Kirsh2009, Capobianco2011, Walsh2011}.  If these
processes strongly shape the final planet distributions, then the
planet distributions from population synthesis models (which truncate
when the gas clears) will form the input to additional simulations
that model post-disk effects and hope to match the presently observed
planet distributions.
 
Current planet formation theory must also adapt to account for remarkable orbital properties of exoplanets.   
Not included here is an analysis of the orbital eccentricities that span the range $e$ = 0--0.93 
\citep[e.g.,][]{Marcy05, Udry2007, Moorhead2011}
and the close-in ``hot Jupiters'' show a wide distribution of inclinations relative to the equatorial plane of the host star 
\citep{Winn2010, Winn2011, Triaud2010, Morton2010}.   
Thus, standard planet formation theory probably requires additional planet-planet gravitational interactions to 
explain these non-circular and non-coplanar orbits \citep{Chatterjee_Ford_Rasio_2010, Wu_Lithwick_2011}.

\subsection{The Future of Kepler}

We strongly advocate for an improved catalog of stellar parameters for 
the $\sim$1000 \Kepler  planet host stars and a comparably-sized control sample.  
Our occurrence measurements and their interpretations would be strengthened by an 
improved knowledge of \rstar, \logg, \feh, and \teff.  
The \rstar\ values from the KIC are only known to  35\% (rms) 
which leads to a proportionally large uncertainty in \rp.  
We saw that hot Jupiters have a significantly lower occurrence in the \Kepler sample than 
in RV surveys.  We were unable to test whether this is due to differing 
metallicities of the host stars because \feh\ is poorly measured in the KIC.  
Similarly, we are unable to completely rule out a metallicity gradient with height above the galactic plane 
as the underlying cause of the observed seven-fold decrease in the occurrence of 2--4 \rearth planets 
with increasing \teff.  

Finally, we  note that Figure \ref{fig:light_curves} shows representative planets 
having $\rp \sim 2.5$ \rearth and $P <  50$ days,
all of which reach SNR $\sim$ 20 in four quarters of \Kepler photometry (and SNR $\sim$ 10 in one quarter).
If we consider the SNR for planets of radius 1 \rearthe, the transit depth is 6 times shallower, 
implying total SNR values near SNR = 20 / 6 = 3.3.   
Thus, planets of 1 \rearthe, even in short periods under 50 days,
would not reach the threshold SNR for meriting a secure detection with
current data in hand.   
For planets of 1 \rearth to reach SNR $\sim$ 6.6, \Kepler must acquire four times more data, i.e.\ five years total, 
still constituting a marginal detection.  
Clearly an extended mission of an additional $\sim$3 yr is needed to bring planets of 1 \rearth to SNR $>$ 7.

\acknowledgements 

We thank E.\,Chiang and H.\,Knuston for helpful conversations.
We gratefully acknowledge D.\,Monet and many other members of the \Kepler team.  
We thank the W.\,M.\,Keck Observatory, and both NASA and the University of California for use of the Keck telescope.  
We are grateful to the Keck technical staff, especially S.\ Dahm, H.\ Tran, and G.\ Hill for support of Keck instrumentation and 
R.\ Kibrick, G.\ Wirth, R.\ Goodrich for support of remote observing.
We extend special thanks to those of Hawai'ian ancestry on whose sacred mountain of Mauna Kea we are privileged to be guests.  
G.\ M. acknowledges NASA grant NNX06AH52G.  
JC-D acknowledges support from The National Center for Atmospheric Research which is sponsored by the National Science Foundation. 
Funding for the \Kepler Discovery mission is provided by NASA's Science Mission Directorate.

\bibliographystyle{apj}
\bibliography{kepocc}

\enddocument